\documentclass[12pt]{article}

\usepackage{a4wide,amsmath,amsthm,amsfonts,amscd,amssymb,eucal,bbm,mathrsfs}

\def\RR{{\mathbb R}}
\def\CC{{\mathbb C}}
\def\NN{{\mathbb N}}
\def\ZZ{{\mathbb Z}}
\def\QQ{{\mathbb Q}}

\def\A{{\mathcal A}}
\def\B{{\mathcal B}}

\def\D{{\mathcal D}}

\def\F{{\mathcal F}}
\def\H{{\mathcal H}}
\def\I{{\mathcal I}}

\def\M{{\mathcal M}}

\def\Z{{\mathcal Z}}

\def\a{\alpha}
\def\b{\beta}
\def\d{\delta}
\def\e{\varepsilon}
\def\f{\varphi}
\def\g{\gamma}
\def\G{\Gamma}

\def\l{\lambda}

\def\p{\psi}

\def\r{\rho}
\def\s{\sigma}

\def\t{\tau}

\def\x{\xi}

\def\gA{\mathfrak A}
\def\gB{\mathfrak B}

\def\Ar{{\gA_\mathrm{r}}}
\def\Br{{\gB_\mathrm{r}}}
\def\gK{\mathfrak K}
\def\gm{\mathfrak m}

\def\isom{\cong}

\def\Ad{{\hbox{\rm Ad\,}}}

\def\id{{\rm id}}
\def\1{{\mathbbm 1}}
\def\Exp{{\rm Exp}}

\def\Tr{\mathrm{Tr}}

\def\geo{\varphi_{\rm geo}}
\def\u1{U(1)}
\def\intervals{\mathcal{I}}
\def\diff{{\rm Diff}}

\def\diffs1{\diff(S^1)}
\def\vir{{\rm Vir}}

\def\supp{{\rm supp}}
\def\psl2r{{\rm PSL}(2,\RR)}
\def\<{\langle}
\def\>{\rangle}
\def\D{C^\infty_c (\RR,\RR)}
\def\Dfin{D_{\mathrm{fin}}}
\def\Dinf{D_{\infty}}

\newtheorem{theorem}{Theorem}[section]

\newtheorem{corollary}[theorem]{Corollary}
\newtheorem{proposition}[theorem]{Proposition}
\newtheorem{lemma}[theorem]{Lemma}
\theoremstyle{remark}
\newtheorem{rem}[theorem]{Remark}

\begin{document}
\date{}

\title{\huge{Thermal States in Conformal QFT. II}}

\author{\textsc{\normalsize Paolo Camassa, Roberto Longo, Yoh Tanimoto, Mih\'aly
Weiner\footnote{Permanent address: Budapest University of Technology and Economics
Department of Analyses, Pf. 91, 1521 Budapest, Hungary}}\\  {\normalsize Universit\`a di Roma ``Tor Vergata'',
Dipartimento di Matematica}
\\  {\normalsize Via della Ricerca Scientifica, 1 - 00133 Roma, Italy}}

\maketitle
\begin{abstract}
We continue the analysis of the set of locally normal KMS states w.r.t.\! the translation group for a local conformal net $\A$ of von Neumann algebras on $\mathbb R$. In the first part we have proved the uniqueness of KMS state on every completely rational net.
In this second part, we exhibit several (non-rational) conformal nets which admit
continuously many primary KMS states.
We give a complete classification of the KMS states on the $U(1)$-current net and on the Virasoro net $\vir_1$ with the central charge $c=1$, whilst for the Virasoro net $\vir_c$ with $c > 1$ we exhibit a (possibly incomplete) list of continuously many primary KMS states.
To this end, we provide a variation of the Araki-Haag-Kastler-Takesaki theorem within the locally normal system framework: if there is an inclusion of split nets $\A\subset \B$ and $\A$ is the fixed point of $\B$ w.r.t.\! a compact gauge group, then any locally normal, primary KMS state on $\A$ extends to a locally normal, primary state on $\B$, KMS w.r.t.\! a perturbed translation.
Concerning the non-local case, we show that the free Fermi model admits a unique KMS state.
\end{abstract}

\begin{center}
{\it
Dedicated to Rudolf Haag on the occasion of his 90th birthday
}
\end{center}

\vskip 3.5cm

\noindent{\footnotesize Research supported in part by the ERC Advanced Grant 227458
OACFT ``Operator Algebras and Conformal Field Theory", PRIN-MIUR, GNAMPA-INDAM and EU network ``Noncommutative Geometry" MRTN-CT-2006-0031962.}

\vskip 0.5cm

\noindent{\footnotesize Email: camassa@mat.uniroma2.it, longo@mat.uniroma2.it,
tanimoto@mat.uniroma2.it, mweiner@renyi.hu}
\newpage

\section{Introduction}\label{introduction}
We continue here our study of the thermal state structure in Conformal Quantum Field Theory, namely we study the set of locally normal KMS states on a local conformal net of von Neumann algebras on the real line with respect to the translation automorphism group.

As is known, local conformal nets may be divided in two classes \cite{LX} that reflect the sector (equivalence class of representations on the circle) structure. For a local conformal net $\A$, to be completely rational (this condition is
characterized intrinsically by the finiteness of the global index \cite{KLM}) is equivalent to the requirement that $\A$ has
only finitely many inequivalent irreducible sectors and all of them have finite index. If $\A$ is not completely rational then either $\A$ has uncountably many inequivalent irreducible sectors or $\A$ has at least one irreducible sector with infinite index.

Faithful KMS states of $\A$ w.r.t. translations are locally normal on the real line (assuming the general split property)
and are associated with locally normal GNS representations of the restriction of the net $\A$ to the real line.
One may wonder whether the structure of these representations, i.e. of the KMS states, also strikingly depends on the
rational/non-rational alternative.

In the first part of our work \cite{CLTW} we have indeed shown the general result that, if $\A$ is a completely rational local conformal net, then there exists only one locally normal KMS state with respect to translations on $\A$ at any fixed inverse temperature $\beta >0$. This state is the geometric KMS state $\f_{\rm  geo}$ which is canonically constructed for any (rational or non-rational) local conformal (diffeomorphism covariant) net.

In this paper we examine the situation when $\A$ is not completely rational.
In contrast to the completely rational case, we shall see that there are non-rational nets
with continuously many KMS states.

We shall focus our attention on two important models. The first one is the free field, i.e.
the net generated by the $\u1$-current. In this model we manage to classify all KMS states.
We shall show that the primary (locally normal) KMS states of the $\u1$-current net are in one-to-one correspondence with real numbers $q\in\RR$; as we shall see, each state $\f^{q}$ is uniquely and explicitly determined by its value on the current. The geometric KMS state is $\geo=\f^{0}$ and any other primary KMS state is obtained by composition of the geometric one with the
automorphisms $\g_{q}$ of the net (see Section \ref{kmsstatesu1}): $\f^{q}=\geo\circ\g_{q}.$

The second model we study is the Virasoro net $\vir_c$, the net generated by the stress-energy tensor with a given central charge $c$. This net is fundamental and is contained in any local conformal net \cite{KL}. If $c$ is in the discrete series, thus $c<1$, the net $\vir_c$ is completely rational, so there exists a unique KMS state by the first part of our work \cite{CLTW}.

In the case $c=1$ we are able to classify all the KMS states. The primary (locally normal) KMS states of the $\vir_{1}$ net w.r.t.
translations are in one-to-one correspondence with positive real numbers
$\left|q\right|\in\RR^{+}$; each state $\f^{\left|q\right|}$ is
uniquely determined by its value on the stress-energy tensor $T$:
\begin{equation*}
\f^{\left|q\right|}\left(T\left(f\right)\right)=\left(\frac{\pi}{12\beta^{2}}+\frac{q^{2}}{2}\right)\int f\, dx.
\end{equation*}
The geometric KMS state corresponds to $q=0$, because it is the restriction
of the geometric KMS state on the $\u1$-current net, and the corresponding
value of the `energy density' $\frac{\pi}{12\beta^{2}}+\frac{q^{2}}{2}$
is the lowest in the set of the KMS states. We construct these KMS states by composing the geometric state with automorphisms on the larger $\u1$-current net.

We mention that, as a tool here, we adapt the Araki-Haag-Kastler-Takesaki theorem
to locally normal systems with the help of split property.
We show that, if we have an inclusion of split nets with
a conditional expectation, then any extremal invariant state on the smaller
net extends to the larger net. The original theorem will be discussed in detail, since we need an extension of a KMS state on the fixed point
subnet to the whole net. Furthermore, we warn the reader that the original proof of the theorem
appears to be incomplete (see \ref{full-extension}),
yet we are able to give a complete proof for the case of split nets (Corollary \ref{co:conformal-extension}),
which suffices for our purpose.

Then we consider the case $c>1$. In this case we produce a continuous family which is probably exhaustive. While we leave open the problem of the completeness of this family, we mention that the formulae on polynomials of fields should be useful.
There is a set of primary (locally normal) KMS states of the $\vir_{c}$
net with $c>1$ w.r.t. translations in one-to-one correspondence with
positive real numbers $\left|q\right|\in\RR^{+}$; each state $\f^{\left|q\right|}$
can be evaluated on the stress-energy tensor
\begin{equation*}
\f^{\left|q\right|}\left(T\left(f\right)\right)=\left(\frac{\pi}{12\beta^{2}}+\frac{q^{2}}{2}\right)\int f\, dx
\end{equation*}
and the geometric KMS state corresponds to \textup{$q=\frac{1}{\beta}\sqrt{\frac{\pi\left(c-1\right)}{6}}$
and energy density $\frac{\pi c}{12\beta^{2}}$.} It is even possible to evaluate $\f^{\left|q\right|}$
on polynomials of the stress-energy tensor and these values are already determined by
the value above on $T(f)$, hence by the number $|q|$. This should give an important information for the complete classification.

We shall also consider a non-local rational model. We will see that there is only one KMS state at each temperature in the free Fermi model. This model contains the Virasoro net $\vir_c$ with $c=\frac{1}{2}$, which is completely rational \cite{KL}. Then by a direct
application of the results in Part I, we obtain the existence and the uniqueness
of KMS state in this case.

We end this introduction by pointing out that our results are relevant for the construction of Boundary Quantum Field Theory nets on the interior of the Lorentz hyperboloid. As shown in particular in  \cite{LR11}, one gets such a net from any translation KMS state on a conformal net on the real line, so our results directly apply.

\section{Preliminaries}\label{preliminaries}
Here we collect basic notions and technical devices
regarding nets of observables and thermal states.
Although our main result in this paper is the classification
of KMS states on certain conformal nets on $S^1$,
we need to adapt
standard results on $C^*$-dynamical systems to our
locally normal systems. Since these materials can be stated for more
general nets of von Neumann algebras, we first formulate the problems
without referring to the circle.

\subsection{Net of von Neumann algebras on a directed set}
\subsubsection{Axioms and further properties}\label{ss:axioms}
Let $\I$ be a directed set. We always assume that there is a countable subset
$\{I_i\}_{i \in \NN} \subset \I$ with $I_i \prec I_{i+1}$ of indices such that for any index $I$
there is some $i$ such that $I \prec I_i$.
A {\bf net (of von Neumann algebras)} $\A$ on $\I$ assigns a von Neumann algebra $\A(I)$
to each element $I$ of $\I$ and satisfies the following conditions:
\begin{itemize}
\item {\bf (Isotony)} If $I \prec J$ then $\A(I) \subset \A(J)$.
\item {\bf (Covariance)} There is a strongly-continuous unitary representation $U$ of $\RR$
and an order-preserving action of $\RR$ on $\I$ such that
\[
U(t)\A(I)U(t)^* = \A(t\cdot I),
\]
and for any index $I$ and for any compact set $C \Subset \RR$,
there is another index $I_C$ such that $t\cdot I \prec I_C$ for $t \in C$.
\end{itemize}
Since the net $\A$ is directed, it is natural to consider the
norm-closed union of $\{\A(I)\}_{I\in\I}$. We simply denote
\[
\gA = \overline{\bigcup_{I \in \I} \A(I)}^{\|\cdot\|}
\]
and call it the {\bf quasilocal algebra}. Each algebra $\A(I)$
is referred to as a {\bf local algebra}.
If each local algebra is a factor, then we call $\A$ a {\bf net of factors}.
The adjoint action
$\Ad U(t)$ naturally extends to an automorphism of the quasilocal
algebra $\gA$.
We denote by $\t_t$ this action of $\RR$ and call it {\bf translation} (note that in this article $\tau_t$ is
a one-parameter family of automorphisms, although in Part I \cite[Section 2.3]{CLTW}, where we assumed
diffeomorphism covariance, we denoted it by $\Ad U(\tau_t)$ to unify the
notation).

An {\bf automorphism of the net $\A$} (not just of $\gA$) is a family $\{\g_I\}$ of
automorphisms of local algebras $\{\A(I)\}$ such that if $I \prec J$ then
$\g_J|_{\A(I)} = \g_I$. Such an automorphism extends by norm continuity to an automorphism of the
quasilocal algebra $\gA$ which preserves all the local algebras.
Conversely, any automorphism of $\gA$ which preserves each local algebra
can be described as an automorphism of the net $\A$.

A net $\A$ is said to be {\bf asymptotically $\g$-abelian} if there is an
automorphism $\g$ of the quasilocal $C^*$-algebra $\gA$
implemented by a unitary operator $U(\g)$
such that
\begin{itemize}
\item $\g$ is normal on each local algebra $\A(I)$ and maps it
into another local algebra $\A(\g\cdot I)$, where we consider that
the automorphism acts also on the set $\I$ of indices by a little abuse
of notation.
\item for any pair of indices $I,J$ there is a
sufficiently large $n$ such that $\A(I)$ and $\A(\g^n\cdot J) = U(\g)^n\A(J)(U(\g)^{*})^n$
commute,
\item $\g$ and $\t_t$ commute.
\end{itemize}

It is also possible (and in many cases more natural) to consider a one-parameter group $\{\g_s\}$ of automorphisms
for the notion of asymptotic $\g$-abelianness (and weakly $\g$-clustering, see below).
In that case, we assume that $\{\g_s\}$ is implemented by a strongly-continuous
family $\{U(\g_s)\}$ and the corresponding conditions above can be
naturally translated.

We say that a net $\A$ is {\bf split} if, for the countable set $\{I_i\}$ in the definition
of the net, there are type I factors $\{\F_i\}$ such that
$\A(I_i) \subset \F_i \subset \A(I_{i+1})$.
Note that in this case the argument in the appendix of \cite{KLM} applies.

\subsubsection{Examples of nets}
The definition of nets looks quite general, but we have principally
two types of examples in mind.

The first comes from the nets on the circle $S^1$ which we have
studied in Part I. For the readers' convenience, we recall the
axioms. A conformal net $\A$ on $S^1$ is a map from the family of intervals
$\I$ of $S^1$ to the family of von Neumann algebras on $\H$ such that:
\begin{enumerate}
\item[(1)] {\bf Isotony.} If $I_1 \subset I_2$, then $\A(I_1) \subset \A(I_2)$.\label{isotony}
\item[(2)] {\bf Locality.} If $I_1 \cap I_2 = \emptyset$, then $[\A(I_1),\A(I_2)] = 0$.
\item[(3)] {\bf M\"obius covariance.} There exists a strongly continuous unitary
representation $U$ of the M\"obius group $\psl2r$ such that
for any interval $I$ it holds that
\begin{equation*}
U(g)\A(I)U(g)^* = \A(gI), \mbox{ for } g \in \psl2r.
\end{equation*}
\item[(4)]{\bf Positivity of energy.} The generator of the one-parameter subgroup of
rotations in the representation $U$ is positive.
\item[(5)] {\bf Existence of vacuum.} There is a unique (up to a phase) unit vector $\Omega$ in
$\H$ which is invariant under the action of $U$,
and cyclic for $\bigvee_{I \in \intervals} \A(I)$.
\item[(6)] {\bf Conformal covariance.} The representation $U$ extends to a projective unitary
representation of $\diffs1$ such that for any interval $I$ and $x \in \A(I)$ it holds that
\begin{gather*}
U(g)\A(I)U(g)^* = \A(gI), \mbox{ for } g \in \diffs1,\\
U(g)xU(g)^* = x, \mbox{ if } \supp(g) \subset I^\prime.
\end{gather*}
\end{enumerate}
Strictly speaking, a net is a pair $(\A, U)$ of a family of von Neumann algebras $\A$ and
a group representation $U$, yet for simplicity we denote it simply by $\A$.

We identify $S^1$ and the one-point compactification $\RR\cup\{\infty\}$ by
the Cayley transform.
If $\A$ is a conformal net on $S^1$, we consider the
restriction $\A|_\RR$ with the family of all finite intervals
in $\RR$ as the index set. The translations in the present setting are the ordinary translations.
If we take a finite translation as $\g$, this system is
asymptotically $\g$-abelian. To consider split property, we can take the sequence of
intervals $I_n = (-n,n)$. It is known \cite{FG} that each local algebra
of a conformal net is a (type $\mathrm{I\!I\!I}_1$) \emph{factor}.
This property is exploited when we extend a KMS state on a smaller net to a larger net.

The second type is a net of observables on Minkowski space $\RR^d$
(see \cite{Haag} for a general account). In this case the index set is
the family of bounded open sets in $\RR^d$. The group of translations in some fixed timelike direction
plays the role of "translations", while a fixed spacelike translation plays the role of $\g$. The net satisfies asymptotic $\g$-abelianness.

In both cases, it is natural to consider the continuous group $\g_s$
of (space-)translations for the notion of $\g$-abelianness.
\subsection{States on a net}
For a $C^*$-algebra $\gA$ and a one-parameter automorphism group
$\{\t_t\}$, it is possible to consider KMS states on $\gA$ with respect to
$\t$. Since our local algebras are von Neumann algebras, it is natural
to consider locally normal objects.
Let $\f$ be a state on the quasilocal algebra $\gA$.
It is said to be
{\bf locally normal} if each restriction of $\f$ to a local
algebra $\A(I)$ is normal.
A {\bf $\b$-KMS state} $\f$ on $\gA$ with respect to $\t$ is a state with the following
properties: for any $x, y \in \gA$, there is an analytic function $f$ in the interior of $D_\b:=\{0 \leq \Im z \leq \b \}$
where $\Im$ means the imaginary part, continuous on $D_\b$, such that
\begin{equation}\label{eq:KMS-condition}
f(t) = \f(x\t_t(y)), f(t+i\b) = \f(\t_t(y)x).
\end{equation}
The parameter $\frac{1}{\b}$ is called the temperature.
In Part I we considered only the case $\b=1$ since our main subject were the conformal nets, in which case
the phase structure is uniform with respect to $\b$. Furthermore, we studied
completely rational models and proved that they admit only one KMS state at each temperature.
Also in this Part II the main examples are conformal, but these models admit continuously many
different KMS states and it should be useful to give concrete formulae which
involve also the temperature.

A KMS state $\f$ is said to be {\bf primary} if the GNS representation of $\gA$ with
respect to $\f$ is factorial, i.e., $\pi_\f(\gA)''$ is a factor.
Any KMS states can be decomposed into primary states \cite[Theorem 4.5]{TW}
in many practical situation, for example if the net is split or if each local algebra
is a factor. Hence, to classify
KMS states of a given system, it is enough to consider the primary ones.

If the net $\A$ comes from a conformal net on $S^1$, namely
if we assume the diffeomorphism covariance, we saw in Part I that
there is at least one KMS state, {\bf the geometric state} $\geo$ \cite[Section 2.8]{CLTW}.
It is easy to obtain a formula for $\geo$ with general temperature $\frac{1}{\b}$.
We exhibit it for later use: let $\omega := \<\Omega,\cdot \Omega\>$ be the vacuum state,
then $\geo := \omega \circ \Exp_\b$, where, for any $I \Subset \RR$,
$\Exp_\b |_{\A(I)} = \Ad U(g_{\b,I}) |_{\A(I)}$ and $g_{\b,I}$ is a diffeomorphism of $\RR$ with compact
support such that for $t\in I$ it holds that $g_{\b,I}(t) = e^{\frac{2\pi t}{\b}}$.

If $\f$ is $\g$-invariant (invariant under an automorphism
$\g$ or a one-parameter group $\{\g_s\}$)
and cannot be written as a linear combination of
different locally normal $\g$-invariant states, then it is said to be
{\bf extremal $\g$-invariant}.

We denote the GNS representation of $\gA$ with respect to
$\f$ by $\pi_\f$, the Hilbert space by $\H_\f$ and the vector which implements the
state $\f$ by $\Omega_\f$. If $\f$ is
invariant under the action of an automorphism $\t_t$ (respectively $\g$, $\g_s$),
we denote by $U_\f(t)$ (resp. $U_\f(\g)$, $U_\f(\g_s)$) the canonical unitary operator which
implements $\t_t$ (resp. $\g$, $\g_s$) and leaves $\Omega_\f$ invariant.
If $\f$ is locally normal, the GNS representation $\pi_\f$ is locally normal
as well,
namely the restriction of $\pi_\f$ to each $\A(I)$ is normal.
Indeed, let us denote the restriction $\f_i := \f|_{\A(I_i)}$.
The representation $\pi_{\f_i}$ is normal on $\A(I_i)$.
The Hilbert space is the increasing union of $\H_{\f_i}$
and the restriction of $\pi_\f$ to $\A(I_i)$ on $\H_{\f_j}$ ($i \le j$)
is $\pi_{\f_j}$, hence is normal. Then $\pi_\f|_{\A(I_i)}$ is normal.

Furthermore, the map $t \mapsto U_\f(t)$ is weakly (and hence strongly)
continuous, since
the one-parameter automorphism $\t_t$ is weakly (or even *-strongly)
continuous and $U_\f(t)$ is defined as the closure of the map
\[
\pi_\f(x)\Omega_\f \longmapsto \pi_\f(\t_t(x))\Omega_\f.
\]
Thus the weak continuity of $t\mapsto U_\f(t)$ follows from the
local normality of $\pi_\f$ and boundedness of $U_\f(t)$, which follows
from the invariance of $\f$. By the same reasoning, if there is
a one-parameter family $\g_s$, the GNS implementation $U_\f(\g_s)$
is weakly continuous.

If for any locally normal $\g$-invariant state $\f$ the algebra $E_0\pi_\f(\gA)E_0$ is abelian, where $E_0$ is the projection onto the space of $U_\f(\g)$-invariant
(resp. $\{U_\f(\g_s)\}$) vectors, then the net $\A$ is said to be {\bf $\g$-abelian}.

A locally normal state $\f$ on $\gA$ is said to be
{\bf weakly $\g$-clustering} if it is $\g$-invariant and
\[
\lim_{N\to \infty}\frac{1}{N}\sum_{n=1}^N \f(\g^n(x)y) = \f(x)\f(y).
\]
for any pair of $x,y \in \gA$. For a one parameter group $\{ \g_s \}$, we define
$\g$-clustering by
\[
\lim_{N\to \infty}\frac{1}{N}\int_0^N \f(\g_s(x)y)ds = \f(x)\f(y).
\]

At the end of this subsection, we remark that, in our principal examples coming
from conformal nets on $S^1$, KMS states are automatically locally normal
by the following general result \cite[Theorem 1]{TW}.
\begin{theorem}[Takesaki-Winnink]\label{local-normality}
Let $\A$ be a net such that $\A(I_i)$ are $\s$-finite properly infinite
von Neumann algebras. Then any KMS-state on $\A$ is locally normal.
\end{theorem}
If $\A$ is a conformal net on $S^1$ defined on a separable Hilbert space,
then each local algebra $\A(I)$ is a type $\mathrm{{I\!I\!I}_1}$ factor,
in particular it is properly infinite, and obviously $\s$-finite,
hence Theorem \ref{local-normality} applies.

\subsection{Subnets and group actions}\label{subnet}
Let $\A$ and $\B$ be two nets with the same index set $\I$
acting on the same Hilbert space.
If for each index $I$ it holds $\A(I) \subset \B(I)$,
then we say that $\A$ is a {\bf subnet} of $\B$ and write simply $\A \subset \B$.
We always assume that each inclusion of algebras has a normal conditional expectation
$E_I: \B(I) \to \A(I)$ such that
\begin{itemize}
\item (Compatibility) For $I \prec J$ it holds that $E_J|_{\B(I)} = E_I$.
\item (Covariance) $\t_t\circ E_I = E_{t\cdot I}\circ \t_t$, and
\end{itemize}
See \cite{LR} for a general theory on nets with a conditional expectation.

Principal examples come again from nets of observables on $S^1$.
As remarked in Part I \cite[Section 2.2]{CLTW}, if we have an inclusion of nets on $S^1$
there is always a compatible and covariant family of expectations.

Another case has a direct relation with one of our main results.
Let $\A$ be a net on $\I$ and assume that there is a family of *-strongly continuous actions $\a_{I,g}$ of a compact Lie group
$G$ on $\A(I)$ such that if $I \subset J$ then $\a_{J,g}|_{\A(I)} = \a_{I,g}$
and $\t_t\circ \a_{I,g} = \a_{t\cdot I,g} \circ \t_t$.
By the first condition (compatibility of $\a$) we can extend $\a$ to an automorphism
of the quasilocal $C^*$-algebra $\gA$, and by the second condition (covariance of $\a$)
$\a$ and $\t$ commute.
Then for each index $I$ we can consider the fixed point
subalgebra $\A(I)^{G}=:\A^G(I)$. Then $\A^G$ is again a net on $\I$. Furthermore,
since the group is compact, there is a unique normalized invariant mean $dg$
on $G$. Then it is easy to see that the map $E(x) := \int_G \a_g(x) dg$
is a locally normal conditional expectation $\A \to \A^G$.
The group $G$ is referred to as the {\bf gauge group} of the inclusion $\A^G \subset \A$.

The *-strong continuity of the group action is valid, for example,
when the group action is implemented by weakly (hence strongly) continuous
unitary representation of $G$. In fact, if $g_n \to g$, then
$U_{g_n} \to U_g$ strongly, hence $\a_{g_n}(x) = \Ad U_{g_n}(x) \to \Ad U_g(x)$
and $\a_{g_n}(x^*) = \Ad U_{g_n}(x^*) \to \Ad U_g(x^*)$
strongly since $\{U_{g_n}\}$ is bounded. This is the case, as are our principal examples,
when the net is defined in the vacuum representation (see \cite[Section 2.1]{CLTW}) and
the vacuum state is invariant under the action of $G$.

If the net $\A$ is asymptotically $\g$-abelian, then we always assume that
$\g$ commutes with $\a_g$.

\subsection{$C^*$-dynamical systems}
A pair of a $C^*$-algebra $\gA$ and a pointwise norm-continuous one-parameter automorphism group $\a_t$
is called a {\bf $C^*$-dynamical system}. The requirement of pointwise norm-continuity is
strong enough to allow extensive general results. Although our main objects are not
$C^*$-dynamical systems, we recall here a standard result.

All notions defined for nets, namely
compact (gauge) group action, asymptotic $\g$-abelianness, $\g$-abelianness,
weakly $\g$-clustering of states and inclusion of systems,
and corresponding results in Section \ref{extension}, except
Corollary \ref{extremal-existence},
have variations for $C^*$-dynamical system (\cite{Kastler}, see also \cite{BR2}).
Among them, important is the theorem of Araki-Haag-Kastler-Takesaki \cite{AHKT}:
for a $C^*$-dynamical system with the fixed point subalgebra with
respect to a gauge group, any KMS state on the smaller
algebra extends to a KMS state with respect to a slightly different
one-parameter group.
In fact, to obtain the full extension, it is necessary to assume that the state $\f$ is faithful
and that there is the net structure. A detailed discussion is collected in
\ref{full-extension}.

\subsection{Regularization}\label{regularization}
To classify the KMS states on $\vir_1$, we need to extend a KMS state
on $\vir_1$ to $\A_{SU(2)_1}$ (explained below). Since $\vir_1$ is
the fixed point subnet of $\A_{SU(2)_1}$ with respect to the action
of $SU(2)$ \cite{Rehren}, one would like to apply Theorem \ref{th:ahkt-precise}.
The trouble is, however, that the theorem applies only to $C^*$-dynamical
systems where the actions of the translation group and the gauge group are pointwise continuous
in norm. The pointwise norm-continuity seems essential in the proof and it is not
straightforward to modify it for locally normal systems; we instead
aim to reduce our cases to $C^*$-dynamical systems.

More precisely, we assume that the net $\A$ has a locally *-strongly continuous action $\t$ of translations (covariance, in subsection \ref{ss:axioms}) and $\a$ of a gauge group $G$ (subsection \ref{subnet}), has an automorphism $\g$ (subsection \ref{ss:axioms}) and they commute, then we construct a $C^*$-dynamical system $(\Ar,\t)$ with the {\bf regular subalgebra} $\Ar$ *-strongly dense in $\gA$.

\begin{proposition}\label{regular-subalgebra}
For any net $\A$ with locally *-strongly continuous action $\t \times \a$ of $\RR \times G$ and an automorphism $\g$ commuting with $\t \times \a$, the set $\Ar$
of elements of the quasilocal algebra $\gA$ on which $\t \times \a$ act pointwise
continuously in norm is a ($\t \times \a,\g$)-globally invariant *-strongly dense $C^*$-subalgebra. Any local element $x \in \A(I)$ can be
approximated *-strongly by a bounded sequence from $\Ar \cap \A(I_C)$ for some $I_C \succ I$.

If we consider a continuous action $\g_s$, then we can take $\Ar$ such that $\Ar$ is
$\{\g_s\}$-invariant and the action of $\g$ is pointwise continuous in norm.
\end{proposition}
\begin{proof}
Let $\Ar$ be the set of elements of $\gA$ on which $\RR\times G$ acts pointwise continuously in norm; $\Ar$ is clearly a $\ast$-algebra and is norm-closed,
hence is a $C^*$-subalgebra of $\gA$. Global invariance follows since $\t$, $\g$ and $\a$ commute.

Let $x$ be an element of some local algebra $\A(I)$.
We consider the smearing of $x$ with a smooth function $f$ on $\RR \times G$ with compact support
\[
x_f := \int_{\RR\times G} f(t,g)\a_g(\t_t(x)) dt dg.
\]
By the definition of net and the compactness of the support of $f$, the integrand belongs to
another local algebra $\A(I_C)$ and the actions $\a$ and $\t$ are normal on $\A(I_C)$,
hence the weak integral can be defined. Smoothness of the actions on $x_f$ is easily
seen from the smoothness of $f$, thus $x_f\in \Ar$.

Take a sequence of functions approximating the Dirac distribution, i.e. a sequence of $f_n$ with $\int_{\RR\times G} f_n(x,g) dxdg = 1$
and whose supports shrink to the unit element in the group $\RR\times G$, then $x_{f_n}$ converges *-strongly to $x$,
since group actions $\a$ and $\t$ are *-strongly continuous by assumption.
Thus, any element $x$ in a local algebra $\A(I)$ can be approximated by a bounded sequence
of smeared elements in a slightly larger local algebra $\A(I_C)$.
As any element in $\gA$ can be approximated in norm (and a fortiori *-strongly) by local elements, $\Ar$ is *-strongly dense in $\gA$.

Moreover, as the actions are norm continuous on $\Ar$, if $x \in \Ar$ then $x_{f_n}$ converges in norm to $x$. This means that the norm closure of the linear space generated by the smeared elements $\{x_f\}$ is an algebra and coincides with $\Ar$.

For a continuous action $\g_s$, it is enough to consider a smearing on $\RR\times G \times \RR$
with respect to the action of $\t \times \a \times \g$.
\end{proof}

\begin{rem}\label{regular-fixed-point}
If $\A$ is the fixed point subnet of $\B$ in the sense of Section \ref{subnet} ($\gA = \gB^G$), then $\Ar = \Br^G$. Indeed, from $\Ar \subset \Br \subset \gB$ it follows that $\Ar \subset \Br^G \subset \gB ^ G=\gA$, since the elements of $\gA$ are $G$-invariant; on the other side, from $\Br^G \subset \gA$ it follows that $\Br^G \subset \Ar$, since the elements of $\Br^G$ are regular. Thus
we obtain an inclusion of $C^*$-dynamical
systems $\Ar \subset \Br$.
\end{rem}

\begin{lemma}\label{clustering-restriction}
If a state $\f$ on the net $\A$ is weakly $\g$-clustering,
then the restriction of $\f$ to the regular system $(\Ar,\t)$ is again weakly $\g$-clustering.
\end{lemma}
\begin{proof}
The definition of weakly $\g$-clustering of a smaller algebra $\Ar$ refers
only to elements in $\Ar$, hence it is weaker than the counterpart for $\gA$.
\end{proof}

\begin{lemma}\label{kms-extension}
Let $\f$ be a locally normal state on $\gA$ which is a KMS state on $\Ar$.
Then $\f$ is a KMS state on $\gA$.
\end{lemma}
\begin{proof}
We only have to confirm the KMS condition for $\gA$.
Let $x,y \in \gA$ and take bounded sequences $\{x_n\}, \{y_n\}$ from $\Ar$
which approximate $x,y$ *-strongly.
Since $\f$ is a KMS state on $\Ar$, there is an analytic function $f_n$
such that
\begin{gather*}
f_n(t) = \f(x_n\t_t(y_n)),\\
f_n(t+i) = \f(\t_t(y_n)x_n).
\end{gather*}
In terms of GNS representation with respect to $\f$, these functions can be
written as
\begin{gather*}
\f(x_n\t_t(y_n)) = \<\pi_\f(x_n^*)\Omega_\f, U_\f(t)\pi_\f(y_n)\Omega_\f\>,\\
\f(\t_t(y_n)x_n) = \<U_\f(t)\pi_\f(y_n^*)\Omega_\f, \pi_\x(x_n)\Omega_\f\>.
\end{gather*}
Note that $\pi_\f(x_n)$ (respectively $\pi_\f(y_n)$) is *-strongly convergent to
$\pi_\f(x)$ (resp. $\pi_\f(y)$) since the sequence $\{x_n\}$ (resp. $\{y_n\}$) is bounded.
Let us denote a common bound of norms by $M$.
We can estimate the difference as follows:
\begin{eqnarray*}
\left|\f(x\t_t(y)) - \f(x_n\t_t(y_n))\right|
&=& \left|\<\pi_\f(x^*)\Omega_\f, U_\f(t)\pi_\f(y)\Omega_\f\>
- \<\pi_\f(x_n^*)\Omega_\f, U_\f(t)\pi_\f(y_n)\Omega_\f\>\right|\\
&\le& M\left\|\pi_\f(x^*) - \pi_\f(x_n^*)\Omega_\f\right\|
+ M\left\|\pi_\f(y) - \pi_\f(y_n)\Omega_\f\right\|
\end{eqnarray*}
and this converges to $0$ uniformly with respect to $t$.
Analogously we see that
$\f(\t_t(y_n)x_n)$ converges to $\f(\t_t(y_n)x_n)$ uniformly.
Then by the three-line theorem (which can be applied because $f_n$ are bounded:
see [3, Prop. 5.3.7]) $f_n(z)$ is uniformly convergent on
the strip $0 \le \Im z \le 1$ and the limit $f$ is an analytic function.
Obviously $f$ connects $\f(x\t_t(y))$ and $\f(t_t(y)x)$, hence
$\f$ satisfies the KMS condition for $\gA$.
\end{proof}

\begin{lemma}\label{kms-faithfulness}
Let $\f$ be a locally normal state on $\gA$ which is a KMS state on $\Ar$.
If each local algebra $\A(I)$ is a factor, then $\f$ is faithful on $\gA$.
\end{lemma}
\begin{proof}
By Lemma \ref{kms-extension}, $\f$ is a KMS state on $\gA$.
The GNS representation $\pi_\f$ is locally normal and hence locally faithful
since each local algebra is a factor, then it is faithful also on the norm
closure $\gA$.
On the other hand, \cite[Corollary 5.3.9]{BR2}
(which applies also to locally normal systems) tells us that
the GNS vector $\Omega_\f$ is separating for $\pi_\f(\gA)''$, thus
$\f(\cdot) = \<\Omega_\f, \pi_\f(\cdot)\Omega_\f\>$ is faithful.
\end{proof}

\section{Extension results}\label{extension}
\subsection{Extension of clustering states}
In this section we provide variations of standard results on $C^*$-dynamical systems.
Parts of the proofs of Lemma \ref{gamma-abelian} and
Proposition \ref{weakly-clustering} are adaptations of \cite{Kastler}  for the locally normal case,
as we shall see.
In particular, when we consider the one-parameter group $\{\g_s\}$,
we need local normality to assure the weak-continuity of the GNS
implementation $\{U_\f(\g_s)\}$.
For some propositions we need the split property in connection with local normality.
\begin{rem}
If we treat one-parameter group $\{\g_s\}$, in the following
propositions (except for Proposition \ref{extremal-decomposition},
where the corresponding modification shall be explicitly indicated)
it is enough to take just the von Neumann algebra
$(\pi_\f(\gA)\cup \{U_\f(\g_s)\})''$ and to consider invariance under
$\{\g_s\}$ or $\{U_\f(\g_s)\}$ and the corresponding notion of
$\g$-clustering property of states. Since $\{U_\f(\g_s)\}$ is
weakly continuous, we can utilize the mean ergodic theorem in this case
as well.
\end{rem}

The following proposition is known (see e.g.\! \cite{BR2, Kastler}).
\begin{proposition}\label{extremal-invariance}
A state $\f$ is extremal $\g$-invariant if and only if
$(\pi_\f(\gA)\cup \{U_\f(\g)\})'' = B(\H_\f)$.
\end{proposition}
Note that any finite convex decomposition of a locally normal state consists of
locally normal states, because a state dominated by a normal state is normal, too.

The following is essential to our argument of extension for locally normal systems.
\begin{theorem}[\cite{Di82}, A\! 86]\label{extremality-criterion}
Let $\H = \int^\oplus_X \H_\lambda d\mu(\lambda)$ be a
direct integral Hilbert space, $T_i = \int^\oplus_X T_{i,\lambda}\mu(\lambda)$
be a sequence of decomposable operators, $\M$ be the von Neumann algebra generated
by $\{T_i\}$, and $\M_\lambda$ be the von Neumann algebra generated
by $\{T_{i,\lambda}\}$.
Then the algebra $\Z$ of diagonalizable operators is maximally commutative in $\M'$
if and only if $\M_\lambda = B(\H_\lambda)$ for almost all $\lambda$.
\end{theorem}

Since we assume the split property of the net $\A$, there is a sequence of
indices $I_i$ and type I factors $\F_i$.
Let $K_i$ be the ideal of compact operators of $\F_i$,
and $\gK$ be the $C^*$-algebra generated by $\{K_i\}$.
With a slight modification about the index set, the following
applies to our situation.
\begin{theorem}[\cite{KLM}, Proposition 56]\label{klm}
Let $\pi$ be a locally normal representation of a split net $\A$ on a
separable Hilbert space and denote by $\pi_\gK$ the restriction to the algebra
$\gK$. If we have a disintegration
\[
\pi_\gK = \int^\oplus_X \pi_\l d\mu(\l),
\]
then $\pi_\l$ extends to a locally normal representation $\widetilde{\pi}_\l$
of $\gA$ for almost all $\l$.
\end{theorem}

We need further a variation of a standard result. The next Proposition would follow
from a general decomposition of an invariant state into extremal invariant states and
\cite[Corollary 5.3]{TW} which affirms that any decomposition is locally normal. In the
present article we take another way through decomposition of representation.
\begin{proposition}\label{extremal-decomposition}
Let $\f$ be a locally normal  $\g$-invariant state of the $C^*$-algebra $\gA$ and $\pi_\f$ be the corresponding GNS representation,
then $\f$ decomposes into an integral of locally normal extremal $\g$-invariant states.
\end{proposition}
\begin{proof}
We take a separable subalgebra $\gK$ as above analogously as in \cite{KLM}.
We fix a maximally abelian subalgebra $\gm$ in the commutant
$(\pi_\gK(\gK)\cup \{U_\f(\g)\})'$.
Since $\gK$ is separable, we can apply \cite[Theorem 8.4.2]{Di82} to obtain
a measurable space $X$, a standard measure $\mu$ on $X$,
a field of Hilbert spaces $\H_\l$ and a field of representations
$\pi_\l$ such that the original restricted representation $\pi_\gK$ is unitarily
equivalent to the integral representation:
\[
\pi_\gK = \int^\oplus_X \pi_\l d\mu(\l)
\]
and $\gm = L^\infty (X,\mu)$.
Now, by Theorem \ref{klm} (note that the representation space $\H_\f$ of the
GNS representation with respect to a locally normal state $\f$ is separable
since we assume that the original net $\A$ is represented on a separable Hilbert space $\H$),
we may assume that $\pi_\l$ is locally normal
for almost all $\l$, hence it extends to a locally normal
representation $\widetilde{\pi}_\l$ and the original representation
$\pi_\f$ decomposes into
\[
\pi_\f = \int^\oplus_X \widetilde{\pi}_\l d\mu(\l).
\]
Furthermore, the GNS vector $\Omega_\f$ decomposes into a direct integral
\[
\Omega_\f = \int^\oplus_X \Omega_\l d\mu(\l).
\]
The representative $U_\f(\g)$ decomposes into direct integrals as well,
since $\gm$ commutes with $U_\f(\g)$:
\[
U_\f(\g) = \int^\oplus_X U_\l(\g) d\mu(\l).
\]
 From this it holds that $\Omega_\l$ is invariant under $U_\l(\g)$, thus
 the state $\f_\l(\cdot) := \<\Omega_\l, \pi_\l(\cdot)\Omega_\l\>$ is invariant
under the action of $\g$, for almost all $\l$.
By the definition of the direct integral it holds that
\[
\f = \int^\oplus_X \f_\l d\mu(\l).
\]
It is obvious that $\f_\l$ is locally normal.

It remains to show that each $\f_\l$ is extremal $\g$-invariant.
By assumption, $\gm$ is maximally commutative in the commutant of $(\pi_\gK(\gK)\cup \{U_\f(\g)\})''$.
This von Neumann algebra is generated by a countable dense subset $\{\pi_\gK(x_i)\}$ and
a representative $U_\f(\g)$.
Then, by Theorem \ref{extremality-criterion}, this is equivalent to the condition
that $(\{\pi_\lambda(x_i)\}\cup \{U_\lambda(\g)\})'' = B(\H_\lambda)$, namely
$\f_\lambda$ is extremal $\g$-invariant.

If we consider a continuous family $\{\g_s\}$, we only have to take
a countable family of operators
$\{\pi_\gK(x_i)\}\cup \{U_\f(\g_s)\}_{s \in \QQ}$.

\end{proof}

\begin{corollary}\label{extremal-existence}
Let $\A \subset \B$ be an inclusion of split nets with a locally normal conditional expectation
which commutes with $\g$.
If $\f$ is an extremal $\g$-invariant state on $\gA$, then $\f$ extends to
an extremal $\g$-invariant state on the quasilocal algebra $\gB$ of the net $\B$.
\end{corollary}
\begin{proof}
The composition $\f\circ E$ is a $\g$-invariant state on $\gB$.
By Proposition \ref{extremal-decomposition}, $\f\circ E$ can be written as an integral
of extremal $\g$-invariant states:
\[
\f\circ E = \int^\oplus_X \p_\l d\mu(\l).
\]
By assumption, the restriction of $\f\circ E$ to $\gA$ is equal to $\f$,
which is extremal $\g$-invariant, hence the restriction $\p_\l|_\A$ coincides with $\f$
for almost all $\l$.
Hence, each of $\p_\l$ is an
extremal $\g$-invariant extension of $\f$.
\end{proof}

\begin{lemma}\label{gamma-abelian}
If the net $\A$ is asymptotically $\g$-abelian, then it is $\g$-abelian.
\end{lemma}
\begin{proof}
Let $\f$ be a locally normal $\g$-invariant state on $\gA$.
The action of $\g$ is canonically unitarily
implemented by $U_\f(\g)$.
Let $E_0$ be the projection onto the space of $U_\f(\g)$-invariant vectors
in $\H_\f$ and $\Psi_1,\Psi_2 \in E_0\H_\f$.
Let us put $\p(x) = \<\Psi_1,\pi_\f(x)\Psi_2\>$.

By the assumption of asymptotically $\g$-abelianness,
it is easy to see that
\[
\lim_{N\to\infty} \frac{1}{N}\sum_{i=1}^N \p(\g^n(x)y) =
\lim_{N\to\infty} \frac{1}{N}\sum_{i=1}^N \p(y\g^n(x)).
\]
On the other hand, by the mean ergodic theorem we have
\begin{eqnarray*}
\lim_{N\to\infty} \frac{1}{N}\sum_{i=1}^N \p(\g^n(x)y)
&=& \lim_{N\to\infty} \frac{1}{N}\sum_{i=1}^N \<\Psi_1,U_\f(\g)^n\pi_\f(x)(U_\f(\g)^*)^n\pi_\f(y)\Psi_2\> \\
&=& \lim_{N\to\infty} \frac{1}{N}\sum_{i=1}^N \<\Psi_1,\pi_\f(x)(U_\f(\g)^*)^n\pi_\f(y)\Psi_2\> \\
&=& \<\Psi_1,\pi_\f(x)E_0\pi_\f(y)\Psi_2\> \\
&=& \<\Psi_1,E_0\pi_\f(x)E_0\pi_\f(y)E_0\Psi_2\>.
\end{eqnarray*}
Similarly we have $\lim_{N\to\infty} \frac{1}{N}\sum_{i=1}^N \p(y\g^n(x)) = \<\Psi_1,E_0\pi_\f(y)E_0\pi_\f(x)E_0\Psi_2\>$.
Together with the above equality we see that
$\<\Psi_1,E_0\pi_\f(x)E_0\pi_\f(y)E_0\Psi_2\> = \<\Psi_1,E_0\pi_\f(x)E_0\pi_\f(y)E_0\Psi_2\>$,
which means that $E_0\pi_\f(x)E_0$ and $E_0\pi_\f(y)E_0$ commute.
\end{proof}

\begin{proposition}\label{weakly-clustering}
If $\f$ is a locally normal $\g$-invariant state on the asymptotically $\g$-abelian net $\A$,
then the following are equivalent:
\begin{itemize}
\item[(a)] in the GNS representation $\pi_\f$, the space of invariant vectors
under $U_\f(\g)$ is one dimensional.
\item[(b)] $\f$ is weakly $\g$-clustering.
\item[(c)] $\f$ is extremal $\g$-invariant.
\end{itemize}
\end{proposition}
\begin{proof}
First we show the equivalence (a)$\Leftrightarrow$(b).
By the asymptotic $\g$-abelianness we have
\[
\lim_{N\to\infty} \frac{1}{N}\sum_{i=1}^N \f(\g^n(x)y) =
\lim_{N\to\infty} \frac{1}{N}\sum_{i=1}^N \f(y\g^n(x)),
\]
and it holds by the mean ergodic theorem that
\begin{eqnarray*}
\lim_{N\to\infty} \frac{1}{N}\sum_{i=1}^N \f(\g^n(x)y)
&=& \<\Omega_\f,E_0\pi_\f(x)E_0\pi_\f(y)E_0\Omega_\f\>, \\
\lim_{N\to\infty} \frac{1}{N}\sum_{i=1}^N \f(y\g^n(x))
&=& \<\Omega_\f,E_0\pi_\f(y)E_0\pi_\f(x)E_0\Omega_\f\>.
\end{eqnarray*}

Now if $E_0$ is one dimensional, then
it holds that
\[
\<\Omega_\f,E_0\pi_\f(y)E_0\pi_\f(x)E_0\Omega_\f\> =
\<\Omega_\f,\pi_\f(y)\Omega_\f\>\<\Omega_\f, \pi_\f(x)\Omega_\f\> =
\<\Omega_\f,E_0\pi_\f(x)E_0\pi_\f(y)E_0\Omega_\f\>,
\]
and this is weakly $\g$-clustering.

Conversely, if $\A$ is weakly $\g$-clustering, the
above equality holds and it implies that $E_0$ is one dimensional,
since $\Omega_\f$ is cyclic for $\pi_\f(\gA)$.

Next we see the implication (a)$\Rightarrow$(c).
Let us take a projection $P$ in the commutant $(\pi_\f(\gA)\cup\{U_\f(\g)\})'$.
Since $P$ commutes with $U_\f(\g)$, $P\Omega_\f$ is again an invariant vector.
By assumption the space of invariant vector is one dimensional,
it holds that $P\Omega_\f = \Omega_\f$ or that $P\Omega_\f = 0$.
We may assume that $P\Omega_\f = \Omega_\f$ (otherwise consider $\1-P$).
By the cyclicity of $\Omega_\f$ for $\pi_\f(\gA)$, it is separating for $\pi_\f(\gA)^\prime$, thus $P=\1$.

Finally, we prove the implication (c)$\Rightarrow$(a).
By Lemma \ref{gamma-abelian}, the algebra $E_0\pi_\f(\gA)E_0$ is
abelian, but by assumption (c), $\pi_\f(\gA) \cup \{U_\f(\g)\}$ act
irreducibly and $U_\f(\g)$ acts trivially on $E_0$.
Hence $E_0\pi_\f(\gA)E_0$ acts irreducibly on $E_0$.
This is possible only if $E_0$ is one dimensional.
\end{proof}

\subsection{Extension of KMS states}\label{extension-kms}
In this section we partly follow the steps in \cite{AHKT}.
We give an overview of the proof of Theorem II.4 of \cite{AHKT}
in \ref{full-extension}, where some notations are introduced.

Let $\A \subset \B$ be an inclusion of asymptotically $\g$-abelian split nets
of factors, and suppose that $\A$ is the fixed point subnet
of a locally normal action $\a$ by a separable compact group $G$ which commutes with
$\g$ and $\t$.
We take a weakly $\g$-clustering primary $\t$-KMS state $\f$ on $\A$
and fix a $\g$-clustering extension $\p$ to $\B$ (whose existence is assured
by Corollary \ref{extremal-existence}).

\begin{lemma}\label{lm:extension-to-bg}
There is a one-parameter group $\e_t \in Z(G_\p,G)$ such that
the restriction of $\p$ to $\B^{G_\p}$ is a faithful KMS state with respect
to $\t^\prime_t := \t_t\circ\a_{\e_t}$.
\end{lemma}
\begin{proof}
We consider the inclusion of $C^*$-algebras $\Ar = \Br^G \subset \Br$ and
the restriction of $\t$. This is an inclusion of $C^*$-systems.
The restriction
of $\p$ to the regular subalgebra $\Br$ is still $\g$-clustering
by Lemma \ref{clustering-restriction}.
We claim that the restriction of $\p$ (hence of $\f$) to $\Br^G$ (see Remark \ref{regular-fixed-point})
is still a primary KMS state. Indeed, the GNS representation of $\f|_{\Br^G}$ can be identified
with a subspace of the representation $\pi_\f$ of $\A$. By the local normality,
this subspace for $\Br^G$ (which coincides with $\Ar$) includes the subspace generated by $\A(I)$ for each fixed
index set. The whole representation space of $\pi_\f$ is the closed union of such subspaces,
hence these spaces coincide. Furthermore, by the local normality, $\pi_\f(\Br^G)''$ contains
$\pi_\f(\A(I))''$ for each $I$. Hence the von Neumann algebras generated by $\pi_\f(\A)$ and
$\pi_\f(\Br^G)$ coincide and $\f|_{\Br^G}$ is primary.

Now we can apply Lemma \ref{lm:ahkt-step1} to obtain a one-parameter group
$\e_t \in Z(G_\p,G)$ such that $\p$ restricted to $\Br^{G_\p}$ is
a KMS state with respect to $\t^\prime_t$. Then by Lemmas \ref{kms-extension},
\ref{kms-faithfulness} we see that $\p$ is a KMS state on the net $\B^{G_\p}$
and it is faithful.
\end{proof}

\begin{theorem}\label{thm:variation-Araki-Haag-Kastler-Takesaki}
Let $\A \subset \B$ be an inclusion of asymptotically $\g$-abelian split nets of factors,
and suppose that $\A$ is the fixed point subnet
of a locally normal action $\a$ by a separable compact group $G$ which commutes with
$\g$ and $\t$.
Then, for any weakly $\g$-clustering extension $\p$ to $\B$ of a primary $\t$-KMS state $\f$ on $\A$
(such an extension always exists by Corollary \ref{extremal-existence}),
there is a one-parameter subgroup $(\e\circ \zeta)$ in $G$
such that $\p$ is a primary $\widetilde{\t}$-KMS state
where $\widetilde{\t}_t = \t_t\circ\a_{\e_t\circ\zeta_t}$.
The state $\p$ is automatically faithful.
\end{theorem}
\begin{proof}
The restriction of $\p$ to
$\Br$ is primary as we saw in Lemma \ref{lm:extension-to-bg}.

This time we consider the inclusion $\Br^{G_\p} \subset \Br$.
Any locally normal representation of the regularized
algebra extends to a representation of the net on the same Hilbert space,
hence it is faithful on the quasilocal algebra since we treat a net of factors.
Then we can apply Lemma \ref{lm:ahkt-step2} together with
Theorem \ref{th:ahkt-corrected} to see that there is a one-parameter subgroup
$\zeta_t \in G_\p$ such that $\p|_\Br$ is a $\widetilde{\t}$-KMS state
where $\widetilde{\t}_t = \t_t\circ\a_{\e_t\circ\zeta_t}$ and
$\e_t$ is taken from Lemma \ref{lm:extension-to-bg}.
By Lemma \ref{kms-extension} $\p$ is a KMS state on the net $\B$.
Again by local normality, the primarity of $\p|_\Br$ and
$\p$ are equivalent.
The faithfulness is proved
as in Proposition \ref{lm:extension-to-bg}.
\end{proof}

We have the following corollary. Note that the gauge group of a split net
is separable because the underlying Hilbert space is automatically separable.
\begin{corollary}\label{co:conformal-extension}
Let $\A \subset \B$ be an inclusion of split conformal nets on the real line $\RR$,
and suppose that $\A$ is the fixed point subnet
of a locally normal action $\a$ by a  compact group $G$ which commutes with
translations $\t$.
Then, for every weakly $\t$-clustering primary $\t$-KMS state $\f$ on $\A$,
there exists a weakly $\t$-clustering extension $\p$ to $\B$.
For any such an extension $\p$ there is a one-parameter subgroup $(\e\circ \zeta)$ in $G$
such that $\p$ is a primary $\widetilde{\t}$-KMS state
where $\widetilde{\t}_t = \t_t\circ\a_{\e_t\circ\zeta_t}$.
The state $\p$ is automatically faithful.
\end{corollary}

\section{The $\u1$-current model}
From now on, we discuss concrete examples from one-dimensional
Conformal Field Theory.
We recall some constructions regarding the $\u1$-current and discuss
its KMS states for two reasons: being a free field model, it is simple
enough to allow a complete classification of the KMS states, showing
an example of non completely rational model with multiple KMS states;
it is useful in the classification of states for the Virasoro nets,
whose restrictions to $\RR$ are translation-covariant subnets of
the $\u1$-current net.

\subsection{The $\u1$-current model}\label{u1-current-model}

The $\u1$-current model is the chiral component of the derivative
of massless scalar free field in the 2-dimensional Minkowski space
time. See \cite{BMT, Longo08} for detail.

In the $\RR$ picture, the space ${C_{c}^{\infty}(\RR,\RR)}$ can
be completed to a complex Hilbert space (the one-particle space) with the complex scalar product
$(f,g):=\int_{p>0}2p\overline{\widehat{f}(p)}\widehat{g}(p)$, where
$\widehat{f}$ is the Fourier transform of $f$, and the imaginary unit
is given by $\widehat{\textit{I}f}(p):=-i\,\textrm{sgn}(p)\widehat{f}(p)$.
The imaginary part of the scalar product is a symplectic form $\sigma(f,g):=\int_{\RR}fg^{\prime}dx$.
The $\u1$-current algebra $\A_{\u1}$ is the Weyl algebra constructed
on this symplectic space, generated by Weyl operators $W\left(f\right)=e^{iJ\left(f\right)}$
acting on the corresponding Fock space (if $f$ is a real function, $J(f)$ is essentially
self-adjoint on the finite particle-number subspace).
The net structure is given
by $\A_{\u1}(I):=\{W(f):\supp(f)\subset I\}^{\prime\prime}$.
This defines a conformal net on $S^1$ in the sense of Part I.
The current operators satisfy $\left[J(f),J(g)\right]=i\sigma(f,g)$
and the Weyl operators satisfy
\[
W\left(f\right) W\left(g\right) = W\left(f + g\right) \exp\left(-\frac{i}{2} \s(f,g) \right).
\]

Let us briefly discuss the split property of the $U(1)$-current net.
A sufficient condition for the split property for a conformal net on $S^1$
is the {\bf trace class condition}, namely the condition that
the operator $e^{-sL_0}$, where $L_0$ is the generator of the rotation automorphism,
is a trace class operator for each $s>0$ \cite{DLR,BDL}.
The Fock space is spanned by the vectors of the following form
$J(e_{-n_1})J(e_{-n_2})\cdots J(e_{-n_k})\Omega$, where
$e_n(\theta) = e^{i2\pi n\theta}$, $0 \le n_1 \le n_2 \le \cdots \le n_k$, $k\in\NN$,
and all these vectors are linearly independent and eigenvectors of $L_0$ with
eigenvalue $\sum_{i=1}^k n_i$. Hence the dimension of the eigenspace with
eigenvalue $N$ is $p(N)$, the partition number of $N$. There is an asymptotic
estimate of the partition function \cite{Hardy}: $p(n)\sim \frac{1}{4n\sqrt 3}e^{\pi\sqrt{2n/3}}$.
Hence with some constants $C_s,D_s$, we have
\[
\Tr(e^{-sL_0}) = \sum_{n=0}^\infty p(n)e^{-sn} \le \sum_{n=0}^\infty C_se^{-D_sn},
\]
which is finite for a fixed $s>0$. Namely we have the trace class condition,
and the split property.

The Sugawara construction $T:= \frac{1}{2}:J^2:$, using normal ordering,
gives the stress-energy tensor, satisfying the commutation relations:
\begin{equation}
\left[T(f),T(g)\right]=iT(\left[f,g\right])+i\frac{c}{12}\int_{\RR}f^{\prime\prime\prime}g\, dx\label{eq:stress-energy-tensor-commutation}\end{equation}
with $c=1$ and $\left[f,g\right]=fg^{\prime}-gf^{\prime}$.
This is the relation of $\mathrm{Vect}(S^1)$, which is the Lie algebra
of $\diffs1$. This (projective) representation $T$ of $\mathrm{Vect}(S^1)$ integrates
to a (projective) representation $U$ of $\diffs1$.
Furthermore, $T$ and $J$ satisfy the following commutation relations
\begin{equation}\label{eq:commutator-T-J}
\left[T(f),J(g)\right]=iJ(fg^{\prime}).
\end{equation}
Accordingly, $U$ acts on $J$ covariantly:
if $\g$ is a diffeomorphism of $\RR$, then $U(\g)J(f)U(\g)^* = J(f\circ \g^{-1})$ (see
\cite{BS, Ottesen} for details).

\subsection{KMS states of the $\u1$-current model}

\label{kmsstatesu1}

We give here the complete classification of the KMS states of the $\u1$-current
model, first appeared in \cite[Theorem 3.4.11]{Wang}.

\begin{proposition} \label{pro:gauge-automorphisms-u1}
There is a one-parameter group $q\mapsto\g_{q}$ of automorphisms
of $\A_{\u1}|_{\RR}$ commuting with translations, locally
unitarily implementable, such that
\begin{equation}\label{eq:u1current-translation}
\g_{q}\left(W\left(f\right)\right)=e^{iq\int_{\RR}fdx}W\left(f\right).
\end{equation}
\end{proposition}
\begin{proof}
For any $I\Subset\RR,$ let $s_{I}$ be a function in ${C_{c}^{\infty}(\RR,\RR)}$
such that $\forall x\in I$ $s_{I}\left(x\right)=x$; then $\sigma(s_{I},f):=\int_{\RR}fdx$
if $\supp f\subset I$ and therefore
\[
\Ad W\left(q s_{I}\right)\, W\left(f\right) =
e^{-i\sigma(q s_{I},f)} W\left(f\right) = e^{iq\int_{\RR}fdx} W\left(f\right).
\]
Set $\g_{q}|_{\A\left(I\right)}=\Ad W\left(q s_{I}\right)$, this
is a well-defined automorphism, since
$\Ad W\left(q s_{I}\right)|_{\A\left(I\right)} =
\Ad W\left(q s_{J}\right)|_{\A\left(I\right)}$
when $I\subset J$, which can be extended to the norm closure $\gA_{\u1}$
satisfying \eqref{eq:u1current-translation} and commuting with translations
because so is the integral.
\end{proof}

\begin{lemma}\label{le:extracharge-term}
A state $\f$ is a primary KMS state of the $\u1$-current model if and only if so is $\f \circ \g_q$
for one value (and hence all) of $q\in\RR$.
\end{lemma}
\begin{proof}
By a direct application of the KMS condition and the fact
that $\g_{q}$ is an automorphism commuting with translations.
\end{proof}

\begin{theorem}\label{thm:KMS-states-of-u1}
The primary (locally normal) KMS states of the $\u1$-current model
at inverse temperature $\b$
are in one-to-one correspondence with real numbers $q\in\RR$; each state
$\f^{q}$ is uniquely determined by its value on the Weyl operators
\begin{equation}
\f^{q}\left(W\left(f\right)\right)=e^{iq\int f\, dx}\cdot e^{-\frac{1}{4}\left\Vert f\right\Vert _{S_{\beta}}^{2}}\label{eq:KMS-states-on-u1}
\end{equation}
where $\left\Vert f\right\Vert _{S_{\beta}}^{2}=\left(f,S_{\beta}f\right)$
and the operator $S_{\beta}$ is defined by
$\widehat{S_{\beta}f}$$\left(p\right)$:=$\coth\frac{\beta p}{2}\widehat{f}\left(p\right)$.
The geometric KMS state is $\geo=\f^{0}$ and any other primary
KMS state is obtained by composition of the geometric one with the
automorphisms \eqref{eq:u1current-translation}:\[
\f^{q}=\geo\circ\g_{q}.\]
\end{theorem}
\begin{proof}
The algebra of the $\u1$-current model is a Weyl CCR algebra, for
which the general structure of KMS states w.r.t. a Bogoliubov automorphism
is essentially known: see e.g. \cite[Theorem 4.1]{RST} or \cite[Example 5.3.2]{BR2}.
It is however easier to do an explicit and straightforward calculation for the present case.

Let $\f$ be a KMS state and $f,g\in C_{c}^{\infty}\left(\RR,\RR\right)$.
Recall that a product of Weyl operators is again a (scalar multiple of)
Weyl operator, so that the quasilocal $C^*$-algebra is linearly generated by
Weyl operators. Hence the state $\f$ is uniquely determined by its values
on $\{W(f)\}$. Furthermore, under the KMS condition,
the function $t\mapsto F(t)=\f\left( W\left(f\right) W\left(g_t\right) \right)$,
where $g_{t}\left(x\right):=g\left(x-t\right)$, has analytic continuation in the interior of $D_\b:=\{0 \leq \Im z \leq \b \}$, continuous on $D_\b$, satisfying \begin{equation}\label{eq:explicit-KMS-cond}
F(t+i\b) =
e^{-i\s(g_t,f)} F(t) = e^{i\s(f,g_t)} F(t).
\end{equation}
We search for a solution $F_0$ of the form $F_0(z)=\exp{K(z)}$,
where $K$ is analytic in the interior of $D_\b$ and has to satisfy the logarithm of \eqref{eq:explicit-KMS-cond}, $K(t+i\b)=i\s(f,g_t) + K(t)$.
The Fourier transform of $t\mapsto i\sigma\left(f,g_t\right)$ is
$p\mapsto p\overline{\widehat{f}\left(p\right)}\widehat{g}\left(p\right)$,
thus we have a simple equation for the Fourier transform  w.r.t. $t$:
$\exp (-\b p) \widehat{K}\left(p\right) = \widehat{K}\left(p\right) +
p \overline{\widehat{f}\left(p\right)} \widehat{g}\left(p\right) $,
from which
$\widehat{K}\left(p\right) =
\frac{p \overline{\widehat{f}\left(p\right) } \widehat{g}\left(p\right)}{\exp (-\b p) -1}$.
It can be explicitly checked that $F_0$ is a solution of \eqref{eq:explicit-KMS-cond}; any other solution, divided by the never vanishing function $F_0$, has to be constant (w.r.t. $t$) by analyticity.
The general solution can therefore be written as $F(t) =
c(f,g_t) \cdot F_0(t)$, with $c(f,g_t)$ independent of $t$.

To obtain \eqref{eq:KMS-states-on-u1}, notice that
\[
\f\left(W\left(f+g_t\right)\right)=
F\left(t\right)e^{\frac{i}{2}\sigma\left(f,g_t\right)}=
c(f,g_t) \cdot \exp\left[ K\left(t\right) + \frac{i}{2}\sigma\left(f,g_t\right) \right] ,
\]
and $K\left(t\right)+\frac{i}{2}\sigma\left(f_{t},g\right)$
is the Fourier antitransform of
\[
p\overline{\widehat{f}\left(p\right)}\left(\frac{1}{e^{-\beta p}-1}+\frac{1}{2}\right)\widehat{g}\left(p\right)
=-\frac{1}{2}p\overline{\widehat{f}\left(p\right)}
\coth\frac{\beta p}{2}
\widehat{g}\left(p\right)
=-\frac{1}{2}p\overline{\widehat{f}\left(p\right)}\widehat{S_{\beta}g}\left(p\right)
\]
which is given by
\[
-\frac{1}{2}\int e^{itp}p\overline{\widehat{f}\left(p\right)}\widehat{S_{\beta}g}\left(p\right)dp
=-\frac{1}{2}\left(f,S_{\beta}g_t\right)
=-\frac{1}{4}\left(\left\Vert f+g_t\right\Vert _{S_{\beta}}^{2}-\left\Vert f\right\Vert _{S_{\beta}}^{2}-\left\Vert g_t\right\Vert _{S_{\beta}}^{2}\right),
\]
since $(f,S_\b g_t)$ is a real form.
Note that $\left\Vert g_{t}\right\Vert _{S_{\beta}}^{2}$ is independent of $t$.
We finally have the general solution in the form
\[
\f\left(W\left(f+g_t\right)\right)=
 c(f,g_t) \cdot e^{\frac{1}{4}\left(\|f\|^2_{S_\b}+\|g_t\|^2_{S_\b}\right)}
 \cdot e^{-\frac{1}{4}\left\Vert f+g_t\right\Vert _{S_{\beta}}^{2}}.
\]
Note that factors $\f(W(f+g_t))$ and $e^{-\frac{1}{4}\|f+g_t\|^2}$
depend only on the sum $f+g_t$, hence so does the remaining factor:
we define $c(f+g_t) := c(f,g_t)\cdot e^{\frac{1}{4}\left(\|f\|^2_{S_\b}+\|g_t\|^2_{S_\b}\right)}$.
Since $c(f,g_t)$ and $\|g_t\|_{S_\b}$ are independent of $t$, so is $c(f+g_t)$.
As $\overline{\f(W(f))}=\f(W(-f))$, $\overline{c(f)}=c(-f)$.
Now we have
\[
\f(W(f+g_t)) = c(f+g_t)\cdot e^{-\frac{1}{4}\|f+g_t\|^2_{S_\b}},
\]
and we only have to determine $c(f+g_t)$.

Concerning the continuity, we notice that $\left\Vert f \right\Vert _{S_{\beta}} \geq \left\Vert f \right\Vert$, because $\coth p \geq 1$ for any $p \in \RR_+$; the map $f \mapsto W(f)$ is weakly continuous when $C_{c}^{\infty}\left(\RR,\RR\right)$ is given the topology of the (one-particle space) norm
$\left\Vert \cdot \right\Vert$ and a fortiori of the norm
$\left\Vert \cdot \right\Vert _{S_{\beta}}$; being $\f$ a KMS state and locally normal,
$f \mapsto \f(W(f))$ is continuous w.r.t. both norms
and $f \mapsto c(f)=\f(W(f)) \cdot \exp({-\frac{1}{4}\left\Vert f \right\Vert _{S_{\beta}}^2})$
is continuous w.r.t. the norm $\left\Vert \cdot \right\Vert _{S_{\beta}}$; finally, both $\l \mapsto \l f$ and $t \mapsto f_t$
are continuous w.r.t. the $\left\Vert \cdot \right\Vert _{S_{\beta}}$ norm,
thus in particular $\l \mapsto c(\l f)$ (and trivially the constant function $t \mapsto c(f+g_t)$) is continuous.

If we require $\f$ to be primary, it satisfies the clustering property: for $t\rightarrow \infty$
\[
\f( W(f + g_t) ) = \f( W(f) W(g_t) ) \exp\left(\frac{i}{2} \s(f,g_t)\right) \rightarrow \f( W(f) )\f( W(g) )
\]
and thus
\begin{equation}\label{eq:c-multiplicative}
c(f+g) = c(f) \cdot c(g) ,
\end{equation}
because both $\s(f,g_t)$ and $(f,S_\b g_t)$ go to $0$. It follows that $c(0)=1$, $c(-f)=c(f)^{-1}=\overline{c(f)}$ and $|c(f)|=1$. As $\RR \ni \l \mapsto c(\l f)$ is a continuous curve in $\{z\in\CC : |z|=1 \}$, there is a unique functional $\r : C_{c}^{\infty}\left(\RR,\RR\right) \rightarrow \RR$ s.t. $c(f)=\exp(i \r (f))$, $\r (0)=0$ and $\l\mapsto \r(\l f)$ is continuous.

Clearly, \eqref{eq:c-multiplicative} implies $\r(f+g) - \r(f) - \r(g) \in 2\pi \ZZ$;
by continuity of $\l \mapsto \r(\l f+ \l g) - \r(\l f) - \r(\l g)$ and $\r(0)=0$, we get $\r(f+g) = \r(f) + \r(g)$.
Similarly, from \cite[Proposition 6.1.2]{Manuceau} we know that $\r$ has the same continuity property of $c$, i.e. w.r.t. the $\left\Vert \cdot \right\Vert _{S_{\beta}}$ norm; $c(f_t)=c(f)$ implies $\r(f_t) - \r(f) \in 2\pi \ZZ$, but this difference vanishes because $t \mapsto \r(f_t)$ is continuous.
Therefore, $\r$ is a real, translation invariant and linear functional.
According to \cite{Meisters}, any translation invariant linear functional (even without requiring continuity) $\r$ on $C_{c}^{\infty}\left(\RR,\RR\right)$ is of the form $\r (f)=q \int f(x) dx$.
So, if $\f$ is a primary KMS state, it has to be of the form \eqref{eq:KMS-states-on-u1}. Conversely, Lemma \ref{le:extracharge-term} implies that all these states are KMS.

These are regular states (i.e. $\l \mapsto \f(W(\l f))$ is a $C^\infty$ function $\forall f$) and the one point and two points functions are given by
\begin{eqnarray}
\f^{q}\left(J\left(f\right)\right) & = & q\int f\, dx\label{eq:KMS-states-on-u(1)-current}\\
\f^{q}\left(J\left(f\right)J\left(g\right)\right) & = & \frac{1}{2}\Re\left(f,S_{\beta}g\right)+\frac{i}{2}\sigma\left(f,g\right)+q^{2}\int f\, dx\int g\, dx,
\end{eqnarray}
where $\Re$ means the real part.
The geometric KMS state has to coincide with one of those: it is $\f^{0}$.
This can be proved by noticing that, if $\supp f\subset I$ , \[
\geo\left(W\left(\lambda f\right)\right)=\left(\Omega,\Ad{U\left(\gamma_{I,\beta}\right)W\left(\lambda f\right)}\Omega\right)=\left(\Omega,W\left(\lambda f\circ\gamma_{I,\beta}^{-1}\right)\Omega\right)=e^{-\frac{1}{4}\lambda^{2}\left\Vert f\circ\gamma_{I,\beta}^{-1}\right\Vert ^{2}}\]
where the exponent is a quadratic form in $f$, therefore the state
is regular and taking the derivative w.r.t. $\lambda$ we get $\geo\left(J\left(f\right)\right)=0$,
which implies $q=0$ by comparison with \eqref{eq:KMS-states-on-u(1)-current}.
\end{proof}

\begin{rem}
The gauge automorphism $\gamma_{z}$ defined by the map $J\left(f\right)\mapsto-J\left(f\right)$
acts as a change in the sign of $q$: $\f^{q}\circ\gamma_{z}=\f^{-q}$.
\end{rem}

The 'energy density' of a state can be read from the expectation value of
the stress-energy tensor as the constant $c$ in the formula $\f\left(T\left(f\right)\right)=c \int f\, dx$.
Beside its physical interpretation, this formula is also useful to classify the states on the Virasoro net
(see Sections \ref{kms-states-vir-1} and \ref{kms-states-vir-c>1}).
In order to evaluate $\f^{q}\left( T(f) \right)$, we need two technical lemmas.

In the following, $\Dfin:= \mathrm{span} \left\{ \psi=J(f_1)...J(f_n)\Omega : n\in\NN, f_1,\ldots,f_n\in C^\infty(S^1,\RR) \right\}$ is the space of finite number of particles and $\Dinf := \cap_{n\in\NN} D(L_0^n)$ is the common domain of the powers of $L_0$; $D(L_0^n) \supset \Dinf$ and $\Dfin$ are all dense in the vacuum Hilbert space, contain the space of finite energy vectors and are cores for $L_0^n$ (the following Lemma implies also that $\Dinf \supset \Dfin$).

\begin{lemma}[Energy bounds]\label{le:energy-bounds}
Let $P_n\left(J,T,L_0\right)$ be a (noncommutative) polynomial in $L_0$ and some $J(f_i)$ and $T(f_j)$ of total degree $n$, with $f_i,f_j \in C^\infty(S^1,\RR)$, then $\forall \psi \in \Dinf$
\begin{equation}\label{eq:estimate_J_T}
\left\Vert P_n\left(J,T,L_0\right) \psi \right\Vert \leq r_n \left\Vert (\1+L_0)^n \psi \right\Vert ,
\end{equation}
with an appropriate $r_n$ (depending on $\left\{f_k\right\}$ and on $n$ but not on $\psi$).
\end{lemma}
\begin{proof}
The operators $J(f)$ and $T(f)$ satisfy similar bounds
\begin{equation}\label{eq:energy-bounds-J-T}
\left\Vert J(f) \psi \right\Vert \leq c_f \left\Vert (\1+L_0) \psi \right\Vert \;\;\;\;\;\;\;\;
\left\Vert T(f) \psi \right\Vert \leq c_f \left\Vert (\1+L_0) \psi \right\Vert
\end{equation}
for any $\psi\in\Dfin$ with $c_f$ independent of $\psi$ \cite[ineqalities (2.21) and (2.23)]{BS}, and similar commutation relations on $\Dfin$:
$[L_0 , J(f)] = i J(\partial_\theta f)$, $[L_0 , T(f)] = i T(\partial_\theta f)$. Since $\Dfin$ is a core for $L_0$, $\forall \psi \in D(L_0)$, using a sequence $\psi_n \in \Dfin$, s.t. $\psi_n \rightarrow \psi$ and $L_0 \psi_n \rightarrow L_0 \psi$, and the closedness of $J(f)$ and $T(f)$, the  bounds \eqref{eq:energy-bounds-J-T} hold on  $D(L_0) \supset \Dinf$;
hence the commutators hold also on $D_\infty$, using $\forall \psi \in \Dinf$ a sequence $\psi_n \in \Dfin$, s.t. $\psi_n \rightarrow \psi$ and $L_0^2 \psi_n \rightarrow L_0^2 \psi$ (from which $L_0 \psi_n \rightarrow L_0 \psi$).
One sees also that $D_\infty$ is invariant under $J(f)$ and $T(f)$.

We can generalize the inequalities \eqref{eq:energy-bounds-J-T}, which are equivalent to \eqref{eq:estimate_J_T} for $n=1$, to any $n$. Indeed, induction and commutation relations show that on $\Dinf$
\begin{equation}\label{eq:commutator-L_0-J}
(\1+L_0)^{n} J(f) =
 \sum_{0\leq k\leq n} \left( \begin{array}{c} n \\  k \\ \end{array} \right)
 i^k J(\partial_\theta^k f) (\1+L_0)^{n-k}.
\end{equation}
Then, we use induction in the degree of the polynomial to prove \eqref{eq:estimate_J_T}.
Suppose \eqref{eq:estimate_J_T} holds for degree $n$. Any polynomial of degree $n+1$ is a linear combination of polynomials of degree $n$ multiplied from the right by $J(f)$ or $T(f)$ or $L_0$.

First, let us consider $J(f)$.
$\Vert P_n\left(J,T,L_0\right) J(f) \psi \Vert \leq r_{n} \Vert (1+L_0)^{n} J(f) \psi \Vert$ by induction hypothesis
and, applying \eqref{eq:commutator-L_0-J} (notice that $\1+L_0 \geq L_0 , \1$), the last norm is smaller than
$\sum_{0\leq k\leq n} c_k \Vert J(\partial_\theta^k f_n) (\1+L_0)^{n-k} \psi \Vert $
where each term is estimated, using \eqref{eq:energy-bounds-J-T}, by constants times $\Vert (\1+L_0)^{n+1} \psi \Vert$.

Secondly, we consider $T(f)$. With $T$ in place of $J$, equation \eqref{eq:commutator-L_0-J} still holds and the same argument as above applies.

Finally,
$\Vert P_n\left(J,T,L_0\right) L_0 \psi \Vert \leq r_{n} \Vert (1+L_0)^{n} L_0 \psi
\Vert \leq r_{n} \Vert (1+L_0)^{n+1} \psi \Vert$ and thus \eqref{eq:estimate_J_T} holds for degree $n+1$.

\end{proof}

\begin{lemma}\label{le:domain-invariant-for-W(f)}
$\Dinf$ is invariant for the Weyl operator $W(f)=e^{i J(f)}$, $\forall f\in C^\infty(S^1)$,
and the unitary $U(g)$, $\forall g\in\diffs1$, implementing the conformal symmetry.
\end{lemma}
\begin{proof}
The subspace $\Dfin$ is included in $\Dinf$ by \eqref{eq:estimate_J_T}, and it
is invariant under $L_0$, as $[L_0 , J(f)] = i J(\partial_\theta f)$. Using also the commutator $[J(f),J(g)^k] =  i k \s(f,g) J(g)^{k-1}$ (easy consequence of $[J(f),J(g)] =  i \s(f,g)$), we compute $\forall \psi \in \Dfin$
\begin{equation}\label{eq:commutator-L_0-J^n}
\left[ L_0 , J(f)^n \right] \psi =
\left( i n J(f)^{n-1} J(\partial_\theta f) - \frac{n(n-1)}{2} J(f)^{n-2} \s(\partial_\theta f,f) \right) \psi .
\end{equation}
We apply it to the expansion of Weyl operators $W(f)=\sum_k \frac{i^k}{k!} J(f)^k$,
which is absolutely convergent on $\Dfin$ (it is well known that finite particle vectors are analytic for the free field, see e.g. the proof of \cite[Theorem X.41]{RS2}, with the estimate $\left\| J(f)^k \psi \right\| \leq 2^{k/2} \sqrt{(n+k)!} \|f\|^k \| \psi \|$, where $n$ is the number of particles of $\psi$). By the closedness of $L_0$ and
the absolute convergence of $L_0 \sum_k \frac{i^k}{k!} J(f)^k \psi$,
thanks to \eqref{eq:commutator-L_0-J^n}, we conclude that $W(f) \Dfin$
is in the domain of $L_0$. We then easily compute, using the convergent series, the commutation relations
$W(f)^\ast L_0 W(f) = L_0 - J(\partial_\theta f) + \frac{1}{2}\s(\partial_\theta f,f)$ and their powers
\begin{equation}\label{eq:commutator-L_0-W}
W(f)^\ast L_0^n W(f) \psi =
\left( L_0 - J(\partial_\theta f) + \frac{1}{2}\s(\partial_\theta f,f)\right)^n \psi.
\end{equation}
Finally, \eqref{eq:estimate_J_T} applied to the r.h.s.,
which is a polynomial of degree $n$ in $J(\partial_\theta f)$ and $L_0$,  gives
\begin{equation}\label{eq:estimate_W}
\Vert L_0^n W(f) \psi \Vert \leq r \Vert (\1+L_0)^n \psi \Vert
\end{equation}
$\forall \psi \in \Dfin$. As $\Dfin$ contains the space of finite energy vectors
(the vectors of $\Dfin$ where $f_1, \cdots, f_n$ are trigonometric polynomials),
it is dense in $\Dinf$ and is a core for $L_0^n$;
any $\psi \in \Dinf$ is the limit of a sequence $\{\psi_i : i\in\NN\}$ such that $(\1+L_0)^n \psi_i$ is convergent,
thus, by \eqref{eq:estimate_W} and the closedness of $L_0^n W(f)$, $W(f) \Dinf$ is in the domain of $L_0^n$.
We have proved that $W(f)\Dinf \subset \Dinf$; the same is true for $W(f)^{-1}=W(-f)$, thus $W(f)\Dinf = \Dinf$.

A similar argument apply to $U(g)$. First one consider the case where
$g = \exp T(f)$ is contained in a one-parameter group.
We replace \eqref{eq:commutator-L_0-W} with the known transformation property
of the stress-energy tensor ($L_0=T(1)$, where $1$ has to be understood as
the generator of rotations, the constant vector field on the circle;
in the real line picture, it would be the smooth vector field $x\mapsto 1+x^2$) \cite{FH}:
\begin{equation}\label{eq:commutator-L_0-U}
U(g) L_0^n U(g)^\ast =
\left( T(g_\ast 1) + r_g \1 \right)^n
\end{equation}
and then apply \eqref{eq:estimate_J_T}.
For a general diffeomorphism $g$, it is possible to write $g$ as a finite product
of diffeomorphisms contained in one-parameter groups, since $\diffs1$ is
algebraically simple \cite{Epstein, Mather} and the subgroup generated by
one-parameter groups is normal, hence $\diffs1$ itself. Thus we obtained
the claimed invariance for any element $g$.
\end{proof}

\begin{theorem} \label{th:KMS-states-on-stress-energy-tens}
For any primary KMS state $\f^{q}$ (cf. \eqref{eq:KMS-states-on-u1}) the map $t \mapsto \f^{q} \left( e^{itT(f)} \right)$ is $C^{\infty}$, $\forall f \in \D$, and the expectation value of the stress-energy tensor is given by
\begin{equation}
\f^{q}\left(T\left(f\right)\right) =
\left(\frac{\pi}{12\beta^{2}}+\frac{q^{2}}{2}\right)\int f\, dx. \label{eq:KMS-states-on-stress-energy-tens}
\end{equation}
Moreover, in the GNS representation $(\pi_{\f^q},\H_{\f^q},\Omega_{\f^q})$, $\Omega_{\f^q}$ is in the domain of
any (non commutative) polynomial of the stress-energy tensors $\pi_{\f^q}(T(f_k)):=-i \frac{d}{dt} \pi_{\f^q}(e^{i T(f_k)})$, with $f_k \in \D$, $k=1,\dots , n$.
\end{theorem}

\begin{proof}
Fix $f\in \D$ with $\supp f\subset I \Subset \RR$.

We first consider the case $q=0$.
According to the proof of Proposition 2.4 and Theorem 2.5 of Part I \cite{CLTW},
the GNS representation of $\geo$ is the triple $\left( \pi_{\geo}=\Exp_\b , \H_{\Omega} , \Omega \right)$
and there is a $g_{\b,I}\in \D$ s.t. $\Exp_\b|_{\A\left(I\right)}=\Ad U\left(g_{\b,I}\right)$.
It follows that the one parameter group
$t\mapsto\pi_{\geo}\left(e^{it T(f)}\right) = \Ad U\left(g_{\b,I}\right) \left(e^{it T(f)}\right)$
has a generator $\Ad U\left(g_{\b,I}\right) \left(T(f)\right)$ which can be computed:
indeed, \cite[Proposition 3.1]{FH} proves that, in general diffeomorphism covariant nets,
if $g\in\diffs1$ fixes the point $\infty$,
$\Ad U\left(g\right)T\left(f\right) =
T\left(g_{\ast}f\right) + r^{\RR}\left(g,f\right)$,
with $g_{\ast}f\left(x\right)=g^{\prime}\cdot f\left(g^{-1}\left(x\right)\right)$
and $r^{\RR}\left(g,f\right) = \frac{c}{12\pi} \int \sqrt{g^{\prime}\left(x\right)} \frac{d^{2}}{dx^{2}} \frac{f\left(x\right)} {\sqrt{g^{\prime}\left(x\right)}}dx$ with the central charge $c$ set equal to $1$ for the $\u1$ case.
Therefore, with $g_{\b,I}$ in place of $g$, recalling that $g_{\b,I}(t) = e^{\frac{2\pi t}{\b}}$
on the support of $f$, we get
\begin{equation}\label{eq:stress-energy-in-GNS-of-geometric-KMS-state}
\pi_{\geo}\left(T(f)\right) = \Ad U\left(g_I\right)T\left(f\right) =
T\left({g_I}_{\ast}f\right) + \frac{\pi c}{12\beta^{2}} \int f \,dx.
\end{equation}
The vacuum vector $\Omega$ is
in the domain of the operator
\eqref{eq:stress-energy-in-GNS-of-geometric-KMS-state} and any product of such operators;
from $\left( \Omega , T(h) \Omega \right) = 0$ for any $h\in\D$,
we easily compute \eqref{eq:KMS-states-on-stress-energy-tens}. The case $q=0$ is proved.

We now consider the general case for $q$.
In this case the GNS representation is $\left( \pi_{\f^q}=\Exp\circ \g_q , \H_{\Omega} , \Omega \right)$ with $\g_{q}|_{\A\left(I\right)}=\Ad W\left(q s_{I}\right)$ defined in Proposition \ref{pro:gauge-automorphisms-u1}.
The one parameter group $t\mapsto \Ad U\left(g_{I}\right) \circ \Ad W\left(q s_{I}\right) \left(e^{it T(f)}\right)$ has a self-adjoint generator $\Ad U\left(g_{I}\right) \circ \Ad W\left(q s_{I}\right) \left(T(f)\right)$, which has to be computed. According to Lemma \ref{le:domain-invariant-for-W(f)}, for any $\psi \in \Dfin \subset \Dinf$ with a finite number of particles, $\Ad W\left(q s_{I}\right) \left(T(f)\right) \psi$ is well-defined because $\Dinf$ is in the domain of $T(f)$.
Using, as for equation (\ref{eq:commutator-L_0-J^n}), $[J(f),J(g)^k] =  i k \s(f,g) J(g)^{k-1}$ and $[T(f),J(g)] =  i J(f g^\prime)$, we compute $\forall \psi \in \Dfin$ a generalization of (\ref{eq:commutator-L_0-J^n}):
\begin{equation*}\label{eq:commutator-T-J^n}
\left[ T(f) , J(g)^n \right] \psi =
\left( i n J(g)^{n-1} J(f g^\prime) - \frac{n(n-1)}{2} J(g)^{n-2} \s(f g^\prime,g) \right) \psi .
\end{equation*}
We use a similar argument to that following equation (\ref{eq:commutator-L_0-J^n}). The expansion of Weyl operators $W(g)=\sum_k \frac{i^k}{k!} J(g)^k$ is absolutely convergent on $\Dfin$; using the absolute convergence of $L_0 \sum_k \frac{i^k}{k!} J(g)^k \psi$ and the estimate $\left\Vert T(f) \psi \right\Vert \leq c_f \left\Vert (\1+L_0) \psi \right\Vert$, we conclude that also $T(f) \sum_k \frac{i^k}{k!} J(g)^k \psi$ is absolutely convergent and therefore, by the closedness of $T(f)$, $W(g) \Dfin$
is in the domain of $T(f)$. The convergent series lets us compute (cf.\! (\ref{eq:commutator-L_0-W}))
$ W(g)^\ast T(f) W(g) \psi = \left( T(f) - J(f g^\prime) + \frac{1}{2}\s(f g^\prime,g) \right) \psi $.
In the particular case in which $g=-q s_I$, and thus $f g^\prime = -q f$ (recall that $\supp f \subset I$), we obtain
\begin{equation*}\label{eq:commutator-W-T}
\Ad W\left(q s_{I}\right) \left(T(f)\right) =
T(f) + q J(f) + \frac{q^2}{2} \int f dx
\end{equation*}
on the dense set $\Dfin$ and also on $\Dinf$, where both sides are defined.
We can apply $\Ad U\left(g_I\right)$ to this operator, as $\Dinf$ is invariant
for $U\left(g_I\right)$, and taking into account its action on $J(f)$ and $T(f)$, we get
\begin{equation}\label{eq:stress-energy-in-GNS-of-q-KMS-state}
\pi_{\f^q}\left(T(f)\right) =
T\left({g_I}_{\ast}f\right) + \frac{\pi}{12\beta^{2}} \int f dx + q J\left(f\circ g_I^{-1} \right) + \frac{q^2}{2} \int f dx .
\end{equation}
 $\Omega$ is in the domain of the operator \eqref{eq:stress-energy-in-GNS-of-q-KMS-state} and any power of such operators;
 as before, using also $\left( \Omega , J(h) \Omega \right) = 0$ for any $h\in\D$, we easily compute \eqref{eq:KMS-states-on-stress-energy-tens}.
\end{proof}

We finally observe that the thermal completion (defined in Part I \cite{CLTW}), in the case of the
$\u1$-current model, does not give any new net.
\begin{theorem}
The thermal completion of the $\u1$-current net w.r.t. any of its
primary (locally normal) KMS states is unitarily equivalent to the
original net.\end{theorem}
\begin{proof}
In the case of the geometric KMS state, this is the content of Theorem 2.5 in Part I \cite{CLTW}.
The general case follows from the fact that any other
primary KMS state of the $\u1$-current model is obtained by composition
of the geometric one with an automorphism, so that the local algebras
$\hat{\A}_{\f^q}(e^{2\pi t},e^{2\pi s}) := \A_{\f^q}(t,\infty) \cap \A_{\f^q}(s,\infty)'$
do not depend on the value of $q$.
\end{proof}

\section{The case of Virasoro nets}

\subsection{The geometric KMS state of $\vir_{c}$}

The Virasoro nets $\vir{}_{c}$ with $c<1$ are completely rational
\cite[Cor. 3.4]{KL}, so our results in Part I \cite{CLTW} apply and thus they have
a unique KMS state: the geometric state $\geo$. This is not the case
for $c\geq1$. Before going to the classification of the KMS states
of $\vir_{1}$ and a (possibly incomplete) list of KMS states for
the Virasoro net with central charge $c>1$, we characterize the geometric
state for any $c$ \cite[Theorem 3.6.2]{Wang}.

\begin{theorem}
The (primary locally normal) geometric KMS states of the $\vir_{c}$
net w.r.t. translations assume the following value on the stress-energy tensor
\begin{equation}
\geo\left(T\left(f\right)\right)=\left(\frac{\pi c}{12\beta^{2}}\right)\int f\, dx.\label{eq:geo-on-stress-energy-tens}
\end{equation}
\end{theorem}
\begin{proof}
The evaluation of the state on the stress-energy tensor \eqref{eq:geo-on-stress-energy-tens} follows from  \eqref{eq:stress-energy-in-GNS-of-geometric-KMS-state} using the same argument of the proof of Theorem \ref{th:KMS-states-on-stress-energy-tens}.
\end{proof}

\subsection{KMS states of the Virasoro net $\vir_{1}$}\label{kms-states-vir-1}
Recall \cite[Section 2.3]{CLTW} that the Virasoro net $\vir_1$ is defined as
the net generated by the representatives of diffeomorphisms. In fact, it holds
that $\vir_1(I) = \{e^{iT(f)}: \supp(f) \subset I\}''$, since the latter
contains the representatives of one-parameter diffeomorphisms, which form a
normal subgroup of $\diff(I)$ (the group of diffeomorphisms with support in $I$),
then this turns out to be the full group because $\diff(I)$ is algebraically simple
\cite{Epstein, Mather}. The net $\vir_1$ is realized as a subnet of the $U(1)$-current;
we have seen that $e^{-sL_0}$ is trace class, hence $\vir_1$ is
split as well.

The primary (locally normal) KMS states of the $\u1$-current, restricted
to the Virasoro net, give primary (locally normal) KMS states. They
are still primary because primarity for KMS states is equivalent to extremality in
the set of $\tau$-invariant states \cite[Theorem 5.3.32]{BR2}, and this is in turn equivalent
to the clustering property (Proposition \ref{weakly-clustering})
for asymptotically abelian nets; clustering
property is obviously preserved under restriction. We denote these
states $\f^{\left|q\right|}$.
We know their values on the stress-energy tensor \eqref{eq:KMS-states-on-stress-energy-tens}.
Notice that the two different states $\f^{q}$ and $\f^{-q}$
coincide when restricted to $\vir_{1}$. We have thus a family of primary
(locally normal) KMS states classified by a positive number $\left|q\right|\in\RR^{+}$.
We will show that these exhaust the KMS states on $\vir_1$.

An important observation for this purpose is that the $U(1)$-current net and
$\vir_1$ can be viewed as subnets of an even larger net. Namely, let $\B:=\A_{SU(2)_1}$ be
the net generated by the vacuum representation of the loop group $LSU(2)$ at level $1$ \cite{FG},
or by the $SU(2)$-chiral current at level $1$ \cite{Rehren}, on which the compact group
$SU(2)$ acts as inner symmetry (an automorphism of the net which preserves the vacuum state).
This net satisfies the trace class condition by an analogous estimate as for $U(1)$-current net
in Section \ref{u1-current-model}, hence it is split.
It has been shown \cite{Rehren} that the Virasoro net $\vir_1$ can be realized as the
fixed point subnet of $\B$ with respect to this inner symmetry.
Moreover, as shown in \cite{Carpi98}, all the subnets of $\B$ are classified
as fixed points w.r.t. the actions of closed subgroups of $SU(2)$
(conjugate subgroups give rise to isomorphic fixed points);
in particular, let $\A_{\u1}$ be the $\u1$-current net,
it is the fixed point $\B^H$ of the net $\B$ w.r.t.
the action of the subgroup $H \simeq S^1$ of rotations around a fixed axis.
Therefore, we have the double inclusion
\[
\vir_1 = \B^{SU(2)} \subset \A_{U(1)} = \B^{H} \subset \A_{SU(2)_1} =: \B,
\]
and a complete classification of the KMS states of the intermediate net $\A_{U(1)}$.
As we are not able to directly extend a $\t$-KMS state on $\vir_1$ to a $\t$-KMS state on $\A_{U(1)}$,
we use an auxiliary extension to $\B$ exploiting the existence of the gauge group $SU(2)$ and Corollary \ref{co:conformal-extension}.

\begin{theorem}
The primary (locally normal) KMS states of the $\vir_{1}$ net w.r.t.
translations are in one-to-one correspondence with positive real numbers
$\left|q\right|\in\RR^{+}$; each state $\f^{\left|q\right|}$
can be evaluated on the stress-energy tensor and it gives
\begin{equation}
\f^{\left|q\right|}\left(T\left(f\right)\right)=\left(\frac{\pi}{12\beta^{2}}+\frac{q^{2}}{2}\right)\int f\, dx.\label{eq:KMS-states-on-stress-energy-tens-c=1}
\end{equation}
\end{theorem}

\begin{proof}
For any $q\in\RR$, the restriction of the KMS state $\f^q$ to the $\vir_1$ subnet gives a  KMS state. The evaluation of the state on the stress-energy tensor \eqref{eq:KMS-states-on-stress-energy-tens-c=1},
depending only on $|q|$, follows again from  \eqref{eq:stress-energy-in-GNS-of-geometric-KMS-state} using the same argument of the proof of Theorem \ref{th:KMS-states-on-stress-energy-tens}.

We have to prove that any primary KMS state of $\vir_1$ arises in this way.
Let $\f$ be a primary KMS state of $\vir_1=\B^{SU(2)}$.
By applying Corollary \ref{co:conformal-extension}, we obtain a locally normal primary (i.e. extremal)
$\t$-invariant extension $\tilde{\f}$ on $\B$, which is a KMS state w.r.t. the one parameter group
$t\mapsto \tilde{\t_t} = \t_t \circ \a_{\e_t\circ\zeta_t}$,
with a suitable one parameter group $t \mapsto \e_t\circ\zeta_t\in SU(2)$.
The image of  $t\mapsto \e_t\circ\zeta_t\in SU(2)$ is a closed subgroup $H \simeq S^1$
since $SU(2)$ has rank $1$ and any one-parameter subgroup forms a maximal torus,
therefore, if we consider the subnet $\A = \B^H$, it is $\tilde{\t}$ invariant and,
as $ \tilde{\t_t}|_{\A}=\t_t|_{\A}$, the state $\tilde{\f}$ is a primary KMS state of $\A$
w.r.t. $\t$. It then follows that the KMS state $\f$ of $\vir_1$ is the restriction of
a KMS state $\tilde{\f}|_\A$ of $\A$, isomorphic to the $\u1$-current net $\A_{\u1}$.
\end{proof}

\begin{rem}
The geometric KMS state corresponds to $q=0$, because it is the restriction
of the geometric KMS state on the $\u1$-current net, and the corresponding
value of the `energy density' $\frac{\pi}{12\beta^{2}}+\frac{q^{2}}{2}$
is the lowest in the set of the KMS states.
\end{rem}

\begin{rem}\label{rem:vir_1}
In contrast to the case of the $\u1$-current net
(Theorem \ref{thm:KMS-states-of-u1}), here the different primary
KMS states are not obtained through composition of the geometric one with automorphisms of the net.

By contradiction, suppose that there were an automorphism $\a$ of the net
such that $\f^{\left|q\right|} = \f \circ \a$ with $q\neq 0$.
The KMS condition for $\f \circ \a$ w.r.t. the one parameter group $t \mapsto \t_t$ is equivalent to the KMS
condition for $\f$ w.r.t. the one parameter group $t \mapsto \a \circ \t_t \circ \a^{-1}$ and,
by the uniqueness of the modular group, $\t_t$ has to coincide with $\a \circ \t_t \circ \a^{-1}$,
i.e. the automorphism of the net commutes with translations. By Proposition 4.2 of Part I \cite{CLTW},
$\a$ cannot preserve the vacuum state and, by Lemma 4.5 of Part I,
there is a continuous family of pairwise non unitarily equivalent automorphisms of $\A|_\RR$ commuting with translations.
By Proposition 4.6 of Part I, there is a continuous family of automorphic sectors of $\A$,
which contradicts the fact, proved in \cite{Carpi03},
that $\vir_1$ can have at most countable sectors with finite statistical dimension.

Recall that in Part I the thermal completion net played a crucial role.
Let $\A_\f(t,s) := \pi_\f(\A(t,s))$ and $\A^d_\f(t,s) := \A_\f(t,\infty)\cap\A_\f(s,\infty)'$.
Putting $\A \equiv \vir_1$ and $\f \equiv \f^{\left|q\right|}$ with $q \neq 0$, we have examples for which
\[
\A_\f(t,s) \neq \A^d_\f(t,s) .
\]
Indeed, if the inclusion $\A_\f(t,s) \subset \A^d_\f(t,s)$ were an equality,
as $\A = \vir_1$ has the split property, Theorem 3.1 of Part I tells that
$\f$ would have to be $\geo\circ\a$.
The observation in the previous paragraph would give a contradiction\footnote{This shows that
the formula (10) in \cite{SW} is incorrect and does not hold in general.}.
\end{rem}

\subsection{KMS states of the Virasoro net $\vir_{c}$ with $c>1$}\label{kms-states-vir-c>1}

Here we show a (possibly incomplete) list of KMS states of the net
$\vir_{c}$ with $c>1$.

The restriction of $\vir_{1}$ to the real line $\RR$ can be embedded
as a subnet of the restriction to $\RR$ of the $\u1$-current net.
One can simply define a new stress-energy tensor \cite[equation (4.6)]{BS}, with $k\in\RR$ and
$f\in\D$
\[ \widetilde{T}\left(f\right):=T\left(f\right)+kJ\left(f^{\prime}\right)\]
 and, using the commutation relations \eqref{eq:stress-energy-tensor-commutation},
calculate that \[
\left[\widetilde{T}\left(f\right),\widetilde{T}\left(g\right)\right]=i\widetilde{T}\left(\left[f,g\right]\right)+i\frac{1+k^{2}}{12}\int_{\RR}f^{\prime\prime\prime}g\, dx.\]
It follows that the net generated by $\widetilde{T}\left(f\right)$
as $\vir_{c}\left(I\right):=\left\{ e^{i\widetilde{T}\left(f\right)}:\supp f\subset I\right\} ^{\prime\prime}$
with $I\Subset\RR$, is the restriction to $\RR$
of the Virasoro net with $c=1+k^{2}>1$ \cite{BS}.
We observe that $\vir_c(I) \subset \A_{\u1}(I)$ for
$I \Subset \RR$. Indeed, we know the locality of $J$ and $T$, hence if $\supp(f) \subset I$,
then $e^{i\widetilde{T}}(f)$ commutes with $W(g)$ with $\supp(g) \subset I'$ by the
Trotter formula. By the Haag duality it holds that $e^{i\widetilde{T}(f)} \in \A_{\u1}(I)$.

The primary (locally normal) KMS states of the $\u1$-current, restricted
again to this Virasoro net, give primary locally normal KMS states,
noticing that $\f^{q}\left(J\left(f^{\prime}\right)\right)=q\int f^{\prime}\, dx=0$:
\[
\f^{\left|q\right|}\left(\widetilde{T}\left(f\right)\right)=\f^{\left|q\right|}\left(T\left(f\right)\right)=\left(\frac{\pi}{12\beta^{2}}+\frac{q^{2}}{2}\right)\int f\, dx;\]
as in the $c=1$ case, the restrictions of $\f^{q}$ and $\f^{-q}$ are
equal. We have thus the following
\begin{theorem}
There is a set of primary (locally normal) KMS states of the $\vir_{c}$
net with $c>1$ w.r.t. translations in one-to-one correspondence with
positive real numbers $\left|q\right|\in\RR^{+}$; each state $\f^{\left|q\right|}$
can be evaluated on polynomials of stress-energy tensor $T(f)$ and on a single $T(f)$
it gives:
\begin{equation}
\f^{\left|q\right|}\left(T\left(f\right)\right)
=\left(\frac{\pi}{12\beta^{2}}+\frac{q^{2}}{2}\right)\int f\, dx.\label{eq:KMS-states-on-stress-energy-tens-c>1}
\end{equation}
The geometric KMS state corresponds to $q=\frac{1}{\beta}\sqrt{\frac{\pi\left(c-1\right)}{6}}$
and energy density $\frac{\pi c}{12\beta^{2}}$.
\end{theorem}
\begin{proof}
As in the case of $\vir_{1}$, the restriction of a primary KMS state of the
$\u1$-current net is a primary KMS state and $\f^{q}=\f^{p}$ if and only if
$q=\pm p$.

The last statement on the geometric KMS state follows by comparison
of \eqref{eq:KMS-states-on-stress-energy-tens-c>1} with \eqref{eq:geo-on-stress-energy-tens}.\end{proof}
\begin{rem}
Unlike the $\vir_{1}$ case, here the geometric KMS state does not
correspond either to $q=0$ or the lowest possible value $\frac{\pi}{12\beta^{2}}$
of the energy density.
\end{rem}

\subsubsection*{An argument toward classification}
We give here an argument that could be useful in the classification of KMS states on Virasoro nets.

Let $\f$ be a primary (locally normal) KMS state on the $\vir_{c}$ net w.r.t. translations and suppose that
$\f \left( \left( T(f_1) \cdots T(f_n) \right)^\ast \left( T(f_1)\cdots T(f_n) \right) \right) < \infty$, $f_1,\cdots,f_n\in \D$. This is the case for all the known KMS states, listed above, although we cannot prove it for a general KMS state. As the state is locally normal, the GNS representation $\pi_\f$ is locally normal (thus a unitary equivalence of type III factors) and can be extended to the stress-energy tensors $T(f)$ ($f\in \D$), which are unbounded operators affiliated to local von Neumann algebras. The above hypothesis is equivalent to the requirement that the GNS vector $\Omega_\f$ is in the domain of any (noncommutative) polynomial of the represented stress-energy tensors $\pi_\f (T(f))$.
We show that the values of the state on polynomials of the stress-energy tensor $\f(T(f_1)\cdots T(f_n))$ are uniquely determined by the value of the state on a single stress-energy tensor $\f \left( T(f) \right)$, for $f\in \D$. This fact seems to determine uniquely the KMS state $\f$, as the net is in some sense generated by such polynomials, however this is not a rigorous statement.

First of all, one can
generalize the KMS condition in order to treat unbounded operators: it is shown in \cite[Prop. 3.5.2]{Wang} that equations \eqref{eq:KMS-condition} hold with $x,y$ possibly unbounded operators affiliated to a local algebra, such that $\Omega_\f$ is in the domain of $\pi_\f(x),\pi_\f(x^\ast),\pi_\f(y)$ and $\pi_\f(y^\ast)$.
Then we show, by induction in $n$, that $\f \left( T(f) \right)$, together with the KMS conditions,
uniquely determines the values $\f \left(T(f_1) ... T(f_n)\right)$.
It is obvious for $n=1$.
It is supposed that $\Omega_\f$ is in the domain of the polynomials of $T(f)$, the value of
$\f \left( \left[ T(f_1) ... T(f_{n-1}) , T(f_n) \right] \right)$ can be computed from the values of $\f$ on polynomials of degree $n-1$, using the commutation relations \eqref{eq:stress-energy-tensor-commutation} which hold on $\Omega_\f$.
According to the KMS condition, there is a function $F(t)= \f \left( T(f_1) ... T(f_{n-1}) \t_t T(f_n) \right)$, continuous and bounded in $D_\b:=\{0 \leq \Im z \leq \b \}$ and analytic in its interior, such that $F(t+i\b)-F(t) = \f \left( \left[ T(f_1) ... T(f_{n-1}) , \t_t T(f_n) \right] \right)$. If $G$ has the same properties, then $F-G$ is continuous in $D_\b$, analytic in its interior and $(F-G)(t+i\b)-(F-G)(t) = 0$, thus $F-G$ can be continued to an analytic bounded functions on $\CC$, which has to be a constant. As $\f$ is primary, the clustering property implies that the constant is $0$: $\lim_{t\rightarrow \infty} F(t) = \f \left( T(f_1) ... T(f_{n-1}) \right) \f \left( T(f_{n}) \right)= \lim_{t\rightarrow \infty} G(t)$. Thus $F$ is uniquely determined and, in particular, $\f \left( T(f_1) ... T(f_{n-1}) T(f_n) \right)$.

If the above argument can be made rigorous, one would get that a KMS state
on the Virasoro net is uniquely determined by the value of the 'energy density',
the constant $k$ appearing in $\f(T(f)) = k \int f(x) dx$ (this in the only possible
expression with translation invariance).
In order to prove that the list in \eqref{eq:KMS-states-on-stress-energy-tens-c>1} is complete,
it would be enough to prove that the set of possible energy density has $\frac{\pi}{12 \b^2}$
as greatest lower bound.

\section{The free fermion model}
In this section we consider the free fermion net and the KMS states
on its quasilocal $C^*$-algebra. For an algebraic treatment of this
model, see \cite{Boeckenhauer,MS}.
In contrast to the $U(1)$-current model,
the free boson model, it turns out to admit a unique KMS state (for each
temperature). The model is not local, but rather graded local. It is still
possible to define a (fermionic) net \cite{CKL}.

The free fermion field $\p$ defined on $S^1$ satisfies
the following Canonical Anticommutation Relation (CAR):
\[
\{\p(z),\p(w)\} = 2\pi i \cdot \d(z-w),
\]
and the Hermitian condition $\p(z)^* = z\p(z)$, or, if we consider the smeared field,
we have
\[
\{\p(f),\p(g)\} = \oint_{S^1} \frac{dz}{2\pi iz} f(z) g(z).
\]
We put the Neveu-Schwarz boundary condition: $\p(ze^{2\pi i})=\p(z)$. Then it is possible
to expand $\p(z)$ in terms of Fourier modes as follows.
\[
\p(z) = \sum_{r\in\ZZ+\frac{1}{2}} b_r z^{-r-\frac{1}{2}}.
\]
The Fourier components satisfy the commutation relation
$\{b_s,b_r\} = \d_{s,-r} \1$, $s,r \in \ZZ + \frac{1}{2}$.

There is a faithful *-representation of this algebra which contains
the lowest weight vector $\Omega$, i.e., $b_s\Omega = 0$ for
$s > 0$ (we omit the symbol for the representation since it is faithful).
This representation is M\"obius covariant \cite[Appendix A]{Boeckenhauer}.
Let $U$ be the unitary representation $U$ of $SL(2,\RR)\isom SU(1,1)$ which
makes $\p$ covariant. It holds that $U(g)\Omega = \Omega$.

Let $P$ be the orthogonal projection onto the space
generated by even polynomials of $\{b_s\}$. It commutes with $U(g)$ and the
unitary operator $\G = 2P - \1$ defines an inner symmetry (an automorphism
which preserves the vacuum state $\<\Omega, \cdot\Omega\>$).

For an interval $I$, we put $\A(I) := \{\p(f): \supp(f) \subset I\}''$. Then $\A$ is a
M\"obius covariant fermi net in the sense of \cite{CKL},
and graded locality is implemented
by $Z$, where $Z := \frac{\1-i\G}{\1-i}$. As a consequence, we have twisted Haag duality: It holds that
$\A(I') = Z\A(I)'Z^*$.
In addition, we have Bisognano-Wichmann property: $\Delta^{it}=U(\Lambda(-2\pi t))$,
where $\Delta^{it}$ is the modular group of $\A(\RR_+)$ with respect to $\Omega$
under the identification of $S^1$ and $\RR\cup\{\infty\}$, and $\Lambda$ is
the unique one-parameter group of $SL(2,\RR)$ which projects to
the dilation subgroup in $PSL(2,\RR)$ under the quotient by $\{\1,-\1\}$ \cite{DLR}.

With $\{b_s\}$ we can construct a representation of the Virasoro algebra with
$c = \frac{1}{2}$ as follows (see \cite{MS}):
\[
L_n := \frac{1}{2}\sum_{s>\frac{n}{2}}
\left(s-\frac{n}{2}\right)b_{-s}b_{n+s}, \mbox{ for }n\ge0,
\]
and $L_{-n} = L_n^*$. For a smooth function $f$ on $S^1$, we can define
the smeared stress-energy tensor $T(f) := \sum_n f_nL_n$, where
$f_n = \oint_{S^1} \frac{dz}{2\pi i}z^{-n-1}f(z)$. The two fields $\p$ and $T$ are
relatively local, namely if $f$ and $g$ have disjoint supports, then
$[\p(f),T(g)] = 0$ ($\p(f)$ is a bounded operator and this holds on a core of $T(g)$).

By the twisted Haag duality, we have $e^{iT(g)} \in \A(I)$ if $\supp(g) \subset I$
(since $\p(f)$ is bounded for a smooth function $f$, there is no problem of
domains). Let us define $\vir_\frac{1}{2}(I) := \{e^{iT(g)}: \supp(g) \subset I\}$.
This Virasoro net $\vir_\frac{1}{2}$ has been studied in \cite{KL} and
it has been shown that $\vir_\frac{1}{2}$ admits a unique nonlocal, relatively local
extension with index $2$. Hence the fermi net $\A$ is the extension.
Furthermore, by the relative locality, $\A$ is diffeomorphism covariant
by an analogous argument as in \cite[Theorem 3.7]{Carpi04}.

We consider the restricted net $\A|_\RR$ on $\RR$ as in Section \ref{ss:axioms},
the quasilocal $C^*$-algebra $\gA$ and translation.
\begin{theorem}
The free fermion net $\A$ admits one and only KMS state at each temperature.
\end{theorem}
\begin{proof}
By the diffeomorphism covariance and Bisognano-Wichmann property, we can
construct the geometric KMS state as in Part I \cite[Section 2.8]{CLTW}
(locality is not necessary).
On the other hand, $\vir_\frac{1}{2}$ is completely rational \cite{KL},
hence it admits a unique KMS state. In this case, we have proved without using locality
\cite[Theorem 4.11]{CLTW} that also the finite index extension $\A$
admits only the geometric KMS state.
\end{proof}

\subsubsection*{Acknowledgement}
We would like to thank the referee for pointing out imprecise statements in \ref{full-extension}.

\appendix
\newcommand{\appsection}[1]{\let\oldthesection\thesection
  \renewcommand{\thesection}{Appendix \oldthesection}
  \section{#1}\let\thesection\oldthesection}

\appsection{On the full extension of a KMS state}\label{full-extension}
In this Appendix we discuss the theorem of Araki-Haag-Kastler-Takesaki \cite{AHKT}.
Let $\gB$ be a $C^*$-algebra, $G$ a compact group acting on $\gB$ and
$\gA = \gB^G$ the fixed point with respect to the action of $G$.
We take a KMS state $\f$ on $\A$ and a weakly $\g$-clustering extension $\p$.
If one looks at the statement carefully, it splits into two parts.
The first part (Theorem II.4) claims that there is a distinguished
subgroup $N_\p$ (depending on
$\p$) of $G$ such that $\p$ is a KMS state on $\gB^{N_\p}$ with respect
to an appropriate one-parameter automorphism group $\widetilde{\t}$.
Then the second part (Remark II.4) says that $N_\p$ is trivial when
$\p$ is faithful on $\gA$, so $\p$ is a KMS state on the whole algebra $\gB$.
We believe that the first part is correct,
but the proof of the second part is missing in the paper and we provide a counterexample at the end of this Appendix.
Hence the extension to the full algebra $\gB$ is not clear
in general\footnote{The
same statement of full extension is found, for example, in \cite[Theorem 5.4.25]{BR2}.
But we think that at least the argument is flawed.
We will give later a counterexample to the argument in \cite{BR2}}.
Here we prove this complete extension with an additional assumption, which can be applied
to the case of nets.

Let $G_\p$ be the group of the stabilizers of $\p$:
\[
G_\p := \{g\in G: \p(\a_g(a)) = \p(a) \mbox{ for all } a \in \gB\}.
\]
The actions $\a$ of $G$, $\t$ of $\RR$ and $\r$ of $\ZZ$ are assumed to be norm-continuous.
We always assume that $G$ is compact, the action of $\g$ on $\gB$ is asymptotically
abelian.
We precisely cite (the relevant part of)\cite[Theorem II.4]{AHKT}
(Note that we changed the notation. In the original literature they use
$\A, \F$ for algebras, $\a$ for the time-translation, $\g$ for the compact group action, $\t$ for the space-translation and $\f$ for the state).
\begin{theorem}[Araki-Haag-Kastler-Takesaki]\label{th:ahkt-precise}
Assume that $G$ is separable. Let $\p$ be a weakly $\g$-clustering state of $\gB$,
whose restriction to $\gA$ is an extremal $(\t_t,\b)$-KMS state. Then there exists a closed
normal subgroup $N_\p$ of $G_\p$, a continuous one-parameter subgroup $\e_t$ of $Z(G_\p,G)$ and
a continuous one-parameter subgroup $\zeta_t$ of $G_\p$ such that the restriction of $\p$
to the fixed point algebra under $N_\p$
\[
\gB^{N_\p} = \{a\in\gB: \a_g(a) = a \mbox{ for all } g\in N_\p\}
\]
is a $(\widetilde{\t}_t,\b)$-KMS state where $\widetilde{\t}_t = \t_t\circ\a_{\e_t\circ\zeta_t}$.
\end{theorem}
We recall that the proof of this Theorem is further split into two parts
(\cite[Theorem II.2, Section II.5 and Section II.6]{AHKT}).
\begin{lemma}\label{lm:ahkt-step1}
Under the hypothesis of Theorem \ref{th:ahkt-precise}, there is a one-parameter
subgroup
\[
\RR \ni t \longmapsto \e_t \in Z(G_\p,G)
\]
such that the restriction of $\p$ to $\gB^{G_\p}$ is an
$(\t^\prime_t,\b)$-KMS state where $\t^\prime_t := \t_t\circ\a_{\e_t}$.
\end{lemma}
\begin{lemma}\label{lm:ahkt-step2}
Under the hypothesis of Theorem \ref{th:ahkt-precise}, there is
a continuous one-parameter subgroup $\zeta_t$ of $G_\p$ such that
the restriction of $\f$ to $\gB^{N_\p}$ is an
$(\widetilde{\t}_t,\b)$-KMS state where $\widetilde{\t}_t := \t_t\circ\a_{\e_t\circ\zeta_t}$.
\end{lemma}
We think both of Lemmas are correct, hence the only task is to show that $N_\p$
trivially acts on $\gB$ under certain conditions. Then let us recall how $N_\p$ is defined.

Consider the space of functions
\[
C_\p(G_\p) := \{f_{a,b}^\p \in C(G_\p): f_{a,b}^\p(g) = \p(a\a_g(b)), a,b\in\gB\}.
\]
It has been shown that the norm closure $\overline{C_\p(G_\p)}$ is a Banach subalgebra
of $C(G_\p)$ \cite[Lemma II.3]{AHKT}, thus the intersection
$\overline{C_\p(G_\p)}\cap\overline{C_\p(G_\p)}^*$ is a $C^*$-subalgebra of
$C(G_\p)$. It is easy to see that this intersection is globally invariant under
left and right translation by $G_\p$ since by definition $\p$ is invariant under $G_\p$,
hence there is a closed normal subgroup $N_\p$
such that $\overline{C_\p(G_\p)}\cap\overline{C_\p(G_\p)}^* \cong C(G_\p/N_\p)$,
where the isomorphism intertwines the natural actions of $G_\p$ \cite[Lemma A.1]{AHKT}.
Explicitly, $N_\p$ is defined as follows:
\[
N_\p := \{g\in G_\p: f(g\cdot) = f(\cdot) \mbox{ for all } f\in \overline{C_\p(G_\p)}\cap\overline{C_\p(G_\p)}^*\}
\]

On the other hand, we can define another normal subgroup $N^\prime_\p$ of $G_\p$:
\[
N_\p^\prime := \{g\in G_\p: f_{a,b}^\p(g\cdot) = f_{a,b}^\p(\cdot) \mbox{ for all } a,b\in \gB\}.
\]
It is easy to see, by uniform approximation, that
\[
N_\p^\prime := \{g\in G_\p: f(g\cdot) = f(\cdot) \mbox{ for all } f\in \overline{C_\p(G_\p)}\}.
\]
Hence, $N_\p^\prime \subset N_\p$. Under a general assumption, $N^\prime_\p$ has a simple interpretation.
\begin{lemma}\label{lm:trivial-n}
Suppose that the GNS representation $\pi_\p$ of $\gB$ is faithful. Then
it holds that $N^\prime_\p = \{g\in G_\p: \a_g(a) = a \mbox{ for all } a\in\gB\}$,
namely $N^\prime_\p$ is the subgroup of the elements acting trivially on $\gB$.
\end{lemma}
\begin{proof}
We show that $N^\prime_\p \subset \{g\in G_\p: \a_g(a) = a \mbox{ for all } a\in\gB\}$,
since the other inclusion is obvious. In the GNS representation, the defining equation
of $N^\prime_\p$ is equivalent to
\[
\<\pi_\p(a^*)\Omega_\p,U_\p(g)\pi_\p(b)\Omega_\p\> = \<\pi_\p(a^*)\Omega_\p, \pi_\p(b)\Omega_\p\>, \mbox{ for all } a,b\in \gB,
\]
which implies that $U_\p(g) = \1$ and $\Ad U_\p(g) = \id$. In particular, we have $\pi_\p(\a_g(a)) = \pi_\p(a)$
for all $a\in \gB$ and, by the assumed faithfulness of $\pi_\p$, we obtain $\a_g(a) = a$.
\end{proof}

\begin{theorem}\label{th:ahkt-corrected}
If the GNS representation of $\gB^{G_\p}$ with respect to (the restriction of) $\p$ is
faithful and if $\pi_\p$ is faithful on $\gB$, then $N_\p$ acts trivially on $\gB$.
\end{theorem}
\begin{proof}
We only have to show that $N_\p = N^\prime_\p$ by Lemma \ref{lm:trivial-n} and the
latter hypothesis.
Under the former assumption, we show that
the intersection $\overline{C_\p(G_\p)}\cap\overline{C_\p(G_\p)}^*$
is equal to $\overline{C_\p(G_\p)}$, then the Theorem follows from the definitions of
$N_\p$ and $N^\prime_\p$.

We remark that the assumption implies that $\p$ is faithful on $\gB$.
Indeed, first the assumption that the GNS representation of $\p$ restricted to
$\gB^{G_\p}$ is faithful implies that $\p$ is faithful on $\gB^{G_\p}$,
since the GNS vector of a KMS state is separating \cite[Corollary 5.3.9]{BR2}.
Now let $x \in \gB$ such that $\p(x^*x) = 0$.
Then, by the definition of $G_\p$, $\p$ is invariant under $G_\p$, thus we have
\[
0 = \int_{G_\p} \p(\a_g(x^*x))dg = \p\left(\int_{G_\p} \a_g(x^*x)dg\right).
\]
But $\int_{G_\p} \a_g(x^*x)dg$ is positive and belongs to $\gB^{G_\p}$, hence
must be zero by the faithfulness of $\p$ on $\gB^{G_\p}$. This is possible only if
$x^*x = 0$ by the continuity of $\a$.

As recalled in \ref{ncha}, for $f_{a,b}^\p \in C_\p(G_\p)$, one can take
its Fourier component $f_{a,b_\chi}^\p$ and the original function $f_{a,b}^\p$
is uniformly approximated by its components. Hence it is enough to consider
irreducible representations.
If $f_{a,b}^\p$ contains $\chi$-component for some $a,b\in\gB$, then
this in particular means that $b_\chi \neq 0$.
By the faithfulness of $\p$ on $\gB$ proved above, one sees that
$\p(b_\chi b_\chi^*) \neq 0$. Since $b_\chi^*$ belongs to the irreducible representation
$\overline{\chi}$, one concludes that the conjugate representation
$\overline{\chi}$ is contained in $\H_\p$.
Then any function in $\overline{C_{\overline{\chi}}(G_\p)}$ (see \ref{ncha})
belongs to $\overline{C_\p(G_\p)}$.

Summing up, the adjoint of each component of $f_{a,b}^\p \in C_\p(G_\p)$ belongs
again to $C_\p(G_\p)$ and each function in $\overline{C_\p(G_\p)}$ is recovered from
its components. This completes the proof of self-adjointness of $\overline{C_\p(G_\p)}$.
\end{proof}

The hypothesis of the Theorem are satisfied not only in our case of conformal nets,
as we see in Section \ref{extension-kms}, but also
in a wide class of models of statistical mechanics where
local algebras are finite dimensional factors $M_n(\CC)$.

\subsubsection*{On the proofs of full-extension in the literature}
As noted before, Theorem \ref{th:ahkt-corrected} without the assumption of faithfulness of
$\pi_\p$ is claimed in \cite{AHKT} without proof.
In \cite[Theorem 5.4.25]{BR2} the theorem of full extension (i.e.\! $N_\p$ acts trivially)
is stated with the assumption of
faithfulness of $\f = \p|_\gA$ on $\gA$. But we think that the proof is not complete.
The argument in \cite{BR2} goes as follows. At the first step, they assume that $\p$ is faithful
on $\gA$ and show that $\p$ is a KMS state on $\gB^{G_\p}$. At the second step, they say that one can
assume that $\p$ is invariant under $G$ and the rest follows.
The point is that, in the first extension, the faithfulness of $\p$ on
$\gB^{G_\p}$ is not automatic. The symmetry of the spectrum of $\pi_\p$ is essential
in the second extension and the faithfulness is used for it.
Here we provide an example which shows that this faithfulness does not hold in general. We do not know whether
the theorem holds without these assumptions.
The same construction gives a counterexample to \cite[Remark II.4]{AHKT}.

We take an auxiliary system $(\gB,\gA,\t,\a,\g)$, where
$\gA = \gB^G$ and take a
KMS state $\f$ on $\gA$ with respect to $\t$ and a $\g$-clustering extension $\p$.
Suppose for simplicity that $\p$ is faithful and
has the whole group as the stabilizer: $G_\p := \{g\in G:\p\circ\a_g = \p\} = G$. We have many
such examples: one can take just the geometric KMS state on
the regularized quasilocal algebra of a conformal net
with a compact group action and the inclusion of the fixed point subnet.

Consider now the field system $(\widehat{\gB},\widehat{\gA},\widehat{\t},\widehat{\a},\widehat{\g})$
where $\widehat{\gB} := \gB\oplus\gB$, $\widehat{G}:=(G\times G)\rtimes\ZZ_{2}$
with $\ZZ_2$ acting on $G\times G$ as the flip, $\widehat{\t}_t := \tau_t\oplus\tau_t$,
the action $\widehat{\a}$ of $(G\times G)\rtimes\ZZ_2$ on $\gB\oplus\gB$
being the action $\a$ of each copy of $G$ on each copy of $\gB$
and the action of $\ZZ_2$ as the flip. The fixed point $\gB^{\widehat{G}}$
is the diagonal algebra $\widehat{\gA}\subset\gA\oplus\gA$, which is isomorphic to $\gA$.
The system $(\widehat{\gB},\widehat{\t})$ is asymptotically abelian as so is $(\gB,\t)$.

Let $\pi_i:\widehat{\gB}\to\gB$ be the projections on a component
and $\p_i:=\p\circ\pi_i.$ The two states $\p_i$
are the two $\g$-clustering extensions of $\f$ on $\widehat{\gB}$ (other
extensions are convex combinations of $\p_1$ and $\p_2$
and are KMS states w.r.t. $\widehat{\t}$).
The stabilizer is in both cases $\widehat{G}_{\p_i}=(G\times G)$,
a normal subgroup of $\widehat{G}$, while the flip exchanges the two states:
$\p_i\circ\a_z=\p_{zi}$ for $z\in\ZZ_2$. The intermediate algebra is $\widehat{\gB}^{\widehat{G}_{\p_i}} = \gA \oplus \gA$
and $\p_i$ is obviously not faithful on it; the faithfulness was assumed implicitly in the second step of the proof in \cite{BR2}.

Let $\pi_\p:\gB\to B(\H_\p)$ be the GNS representation
of $\p$ (faithful as $\p$ is faithful), then $\pi_{\p_i}=\pi_\p\circ\pi_i:\widehat{\gB}\to B(\H_\p)$
is not faithful, although it is true that
$\pi_{\p_i}(\widehat{\gA})^{\prime\prime} = \pi_{\p_i}(\widehat{\gB}^{G_{\p_i}})^{\prime\prime}$.

As $\pi_{\p_i}$ is not faithful, although $\p_i$ is faithful on $\widehat{\gA}$,
we cannot deduce that
$N^\prime_{\p_i} := \{g\in\widehat{G}:\p_i(a\widehat{\a}_g(b)) = \p_i(ab) \mbox{ for all } a,b\in\widehat{\gB}\}$
is trivial nor that it acts trivially. Indeed, $N^\prime_{\p_1}=(1,G,1)$ and
$N^\prime_{\p_{2}}=(G,1,1)$; this is in contradiction with \cite[Remark II.4]{AHKT},
since we show that $N^\prime_{\p_i} = N_{\p_i}$, also in this case. By proceeding as in the third paragraph of the proof of
Theorem \ref{th:ahkt-corrected}, let $b_\chi$ belong to the irreducible representation $\chi$
and $\p_1(b_\chi^*b_\chi) \neq 0$. Then, as $b_\chi$ is of the form $b_\chi = b_1\oplus b_2$,
$\p_1(b_\chi b_\chi^*) = \p(b_1 b_1^*) \neq 0$ by the faithfulness of $\p$,
which implies that $\overline\chi$ is contained in $\H_\p$; the rest follows as
in Theorem \ref{th:ahkt-corrected}.
One sees that $N_{\p_i}$ is a normal subgroup
of $\widehat{G}_{\p_i}$, $\p_i$ and $\pi_{\p_i}$ are
faithful neither on $\widehat{\gB}$ nor on $\widehat{\gB}^{N_{\p_{1}}} = \gB\oplus\gA$.
Moreover, the one parameter group $\widehat{\t}$ w.r.t.
which $\p_i$ is KMS is not uniquely defined, as $\p_{1}$
is not faithful and is KMS w.r.t. $\widehat{\t}\circ\a_{(1,g_t,1)}$
for any $t\mapsto g_t\in G$.

\appsection{Noncommutative harmonic analysis}\label{ncha}
Here we briefly summarize elementary methods to treat actions of a compact group $G$
on a $C^*$-algebra. For the classical facts from the representation theory of compact
groups, we refer to the standard textbooks, for example, \cite{Knapp}.
The classical Peter-Weyl theorem says that any irreducible
representation of $G$ is finite dimensional.
To a finite-dimensional representation one can associate a character $\chi$
in the space $C(G)$ of continuous functions on $G$. On this space
$G$ acts by left and right translations. This becomes a pre-Hilbert space by the inner product
induced by the Haar measure and its completion is denoted by $L^2(G)$. The action by
translation is referred to as the left or right regular representation.
Again the Peter-Weyl theorem states that the left or right regular representation contains
any irreducible representation and the multiplicity is equal to its dimension.
If a function $f$ belongs to an irreducible representation $\chi$ of dimension $n$
of the left (or right) regular representation, then
the images of $f$ under right and left translation of $G \times G$ span the whole $n^2$ dimensional space.
Here we call this subspace $C_\chi(G)$.
Two characters $\chi, \chi^\prime$ are orthogonal iff the corresponding representations are disjoint.
Any unitary representation $U$ can be written as the direct sum of irreducible representations.
The decomposition into classes of inequivalent representations is canonical: for a character
$\chi$ associated to an irreducible representation, the map
\[
\xi \mapsto \xi_\chi = \int_G \overline{\chi(g)} U(g)\xi dg.
\]
is the projection from the representation space onto the direct sum of irreducible subrepresentations of $U$ equivalent to
 the one corresponding to $\chi$.
 It holds that $\xi = \sum_\chi \xi_\chi$ and $\xi_\chi \perp \xi_{\chi'}$
if $\chi$ and $\chi'$ are inequivalent. The above formula is an extension of
the Fourier decomposition.

An action $\a$ of $G$ on a $C^*$-algebra $\gB$ is an infinite dimensional representation of $G$ on a Banach space.
It is still possible to define the Fourier components:
for $a \in \gB$, we put
\[
a_\chi := \int_G \overline{\chi(g)}\a_g(a)dg.
\]
In general the sum $\sum_\chi a_\chi$ is not necessarily norm-convergent.
Now let us assume that there is a $G$-invariant state $\p$. Then in the GNS representation
$(\H_\p, \pi_\p, \Omega_\p)$ there is a unitary representation $U_\p$ which implements
the action $\a$. The components defined for $U_\p$ and $\a$ are compatible: we have
\[
\pi_\p(a_\chi)\Omega_\p = \int_G \overline{\chi(g)}\pi_\p(\a_g(a))\Omega_\p
= \int_G \overline{\chi(g)}U_\p(g)\pi_\p(a)\Omega_\p = (\pi_\p(a)\Omega_\p)_\chi.
\]
From the orthogonality in the representation $U_\p$, one sees that
if $\chi$ and $\chi^\prime$ correspond to two disjoint representations, then
$\p((a_\chi)^*\a_g(b_{\chi^\prime})) = \<\pi_\p(a_\chi)\Omega_\p,U_\p(g)\pi_\p(b_{\chi^\prime}))\Omega_\p\> = 0$.
It is immediate to see that the function $g\mapsto\p(a\a_g(b_\chi)) = \p((a_{\overline{\chi}})\a_g(b))$
belongs to $C_\chi(G)$.
The decomposition of the vector $\pi_\p(b) \Omega_\p = \sum_\chi \pi_\p(b_\chi)\Omega_\p$
converges in norm, hence for the function $f_{a,b}^\p(g) := \p(a\a_g(b))$,
the decomposition $f_{a,b}^\p = \sum_\chi f_{a,b}^\chi$ where
$f_{a,b}^\chi(g) = \<\pi_\p(a^*)\Omega_\p,U_\p(g)\pi_\p(b_\chi)\Omega_\p\>$
converges uniformly in the norm of $C(G)$.

\end{document}